\documentstyle[aps,prl,epsf]{revtex}
\begin{document}

\tighten        

\twocolumn[\hsize\textwidth\columnwidth\hsize  
\csname @twocolumnfalse\endcsname              

\title{{\vskip -2cm}\null\hfill hep-ph/9802378(ORNL-CTP-9802)\\ 
{\vskip 1.0cm}
Effects of Parton Intrinsic Transverse Momentum on 
Photon Production in Hard-Scattering Processes}

\vspace{0.1in} \author{Cheuk-Yin Wong} \address{Physics Division, Oak
Ridge National Laboratory, Oak Ridge, Tennessee, U.S.A.}

\vspace{0.1in} \author{Hui Wang} \address{Physics Division, Oak Ridge
National Laboratory, Oak Ridge, Tennessee, U.S.A. \\ and China
Institute of Atomic Energy, Beijing, China}

\date{\today}
\maketitle

\vspace{0.1in}
\begin{abstract}
We calculate the photon production cross section arising from the hard
scattering of partons in nucleon-nucleon collisions by taking into
account the intrinsic parton transverse momentum distribution and the
next-to-leading-order contributions.  As first pointed out by Owens,
the inclusion of the intrinsic transverse momentum distribution of
partons leads to an enhancement of photon production cross section in
the region of photon transverse momenta of a few GeV/c for
nucleon-nucleon collisions at a center-of-mass energy of a few tens of
GeV. The enhancement increases as $\sqrt{s}$ decreases.  Such an
enhancement is an important consideration in the region of photon
momenta under investigation in high-energy heavy-ion collisions.
\end{abstract}


   ]  

\narrowtext
\section{Introduction}

Currently, high-energy heavy-ion collisions are used to produce
matter under extreme conditions.  One of the objectives is to search
for the quark-gluon plasma (QGP) which is expected to exist at high
temperatures and/or high baryon densities.  (For an introduction to
this field, see \cite{QM96,Won94,Hwa95,Sta92,Har96}.)  Photons arising
from the electromagnetic interactions of the constituents of the
plasma will provide information on the properties of the plasma at the
time of their production.  Since photons are hardly absorbed by the
produced medium, they form a relatively `clean' probe to study the
state of the quark-gluon plasma.  The detection of these photons
during the quark-gluon plasma phase will be of great interest in
probing the state of the quark-gluon plasma, if it is ever produced
\cite{Hwa85,McL85,Kaj86,Kap91,Sri92,WA80,WA98,WA9897,Xio92,Hun96}.

Photons are also produced by many other processes in heavy-ion
reactions. They can originate from the decay of $\pi^0$ and $\eta^0$.
As $\pi^0$ particles are copiously produced in strong interactions
between nucleons, photons originating from the decay of $\pi^0$ are
much more abundant than photons produced by electromagnetic
interactions of the constituents of the quark-gluon plasma.  The
photons from the decay of $\pi^0$ and $\eta^0$ can be subtracted out
by making a direct measurement of their yield, obtained by combining
pairs of photons.  Because of the large number of $\pi^0$ produced,
this subtraction is a laborious task, but much progress has been made
to provide meaningful results after the subtraction of the photons
from the $\pi^0$ and $\eta^0$ backgrounds \cite{WA80,WA98,WA9897}.
Photon measurements obtained after the subtraction of the photons from
meson decays are conveniently called measurements of ``direct
photons''.

Direct photons are produced from the interaction of matter in the QGP
phase, a mixed QGP and hadron phase, a pure hadron gas, and hard QCD
processes\cite{Sta92,Har96}.  Different processes give rise to photons
in different momentum regions.  We are interested in photons from the
quark-gluon plasma which are found predominantly in the region of
photons with low transverse momentum, extending to the region of
intermediate transverse momentum of 1-3 GeV/c.  At photon energies up
to 2 GeV, photons can come from the decay of $\pi^0$ and $\eta^0$
resonances as well as from $\rho, \omega, \eta'$, and $a_1$, and the
interaction of the hadron matter via $\pi \rho \rightarrow \gamma
\rho$ and $\pi \pi \rightarrow \gamma \pi$ reactions (see for example
\cite{Xio92,Hun96,Ste96,Li97}).  One may wish to go to the region of
photon transverse momentum $p_{\gamma T}>$ 2 GeV/c.  If a hot
quark-gluon plasma is formed initially, clear signals of photons from
the plasma could be visible by examining photons with $p_{\gamma T}$
in the range 2 $-$ 3 GeV/c \cite{Kaj86,Kap91,Sri92}.  On the other
hand, photons in this region of transverse momenta are also produced
by the collision of partons of the projectile nucleons with partons of
the target nucleons.  Such a contribution must be subtracted in order
to infer the net photons from the quark-gluon plasma sources.

Recent investigations on photon production by parton collisions in
hadron-hadron collisions include the work by Aurenche, Baier,
Douiri, Fontannaz, and Schiff \cite{Aur87,Aur88} and by Baer, Ohnemus,
and Owens \cite{Bae90}, who have performed QCD calculations up to
second order in $\alpha$.  Extensive comparisons with experimental
data have been carried out covering a large range of incident energies
and photon transverse momenta.  In these QCD calculations, as well as
other similar QCD calculations \cite{Cle95}, the intrinsic transverse
momenta of the partons have been neglected.

Because of the finite size of the transverse dimension of a nucleon,
one expects that the partons in a nucleon have an intrinsic transverse
momentum distribution with a width of the order of
$\sqrt{2}$$\hbar/0.5$ fm.  The intrinsic transverse momentum
distribution affects the distribution of the produced photons.
Previous investigations of the photon transverse momentum distribution
in leading-order ($LO$) calculations with a constant $K$-factor
already indicate the importance of the intrinsic transverse momentum
of the partons \cite{Owe87}.  For $p_{\gamma T}$ from 3 to 8 GeV/c,
the photon production cross sections for hadron-hadron collisions at
at 19.4 GeV $<\sqrt{s}< 63$ GeV are enhanced when the intrinsic
transverse momentum distribution is taken into account, in agreement
with earlier measurements.  These findings are further supported by
recent experimental and theoretical investigations for hadron-proton
collisions at $\sqrt{s}=$31.7 and 38.8 GeV \cite{Apa97}.  The
enhancement increases as $\sqrt{s}$ decreases.  Therefore, in the
region of our interest involving photons with $p_{\gamma T}$ of about
2-3 GeV/c at a center-of-mass energy of about 20 GeV, as in the
measurements of the WA80 and the WA98 Collaborations \cite{WA80,WA98},
the intrinsic transverse momentum of partons plays an important role
and cannot be neglected.  How the photon transverse momentum
distributions at different $\sqrt{s}$ are affected by the parton
intrinsic transverse momentum is the main subject of our
investigation.

There is also another important effect which is important in photon
production.  Previous calculations of photon production indicate the
importance of higher-order QCD terms \cite{Aur87,Aur88,Bae90}.  The
next-to-leading-order calculations lead to a correction factor, the
$K$-factor, with a magnitude of about 2.  It is necessary to include
the effects of the next-to-leading-order corrections.  One expects
that the next-to-leading-order effects are nearly independent of the
intrinsic transverse momentum.  One can include the effects of the
intrinsic transverse momentum and the next-to-leading-order
corrections by first treating them separately and then combining them
as independent multiplicative factors.  The final results can be
compared with experimental data.  Good agreement with experimental
data at different reaction energies will form the basis of
extrapolation to the region of interest in high-energy heavy-ion
collisions.

Previous work by Cleymans $et~al.$\cite{Cle95} on photon production in
$pp$ collisions covers a region from $\sqrt{s}=23$ GeV to 1.8 TeV.
These authors use the next-to-leading-order calculation program of
Aurenche $et~al$\cite{Aur87}, without assuming parton intrinsic
transverse momentum.  However, the results of these authors cannot be
interpreted as the conclusive evidence for the absence of any parton
intrinsic momentum effect at energies of $\sqrt{s}\sim$ 20 GeV,
because their theoretical results are lower than the UA6 data at
$\sqrt{s}=24.3$ GeV by about a factor of 2 (see Figure 3 of
\cite{Cle95}), and the theoretical results are lower than the E706
data at $\sqrt{s}=30.6$ GeV by about a factor of 1.7 (see Figure 4 of
\cite{Cle95}). As mentioned above, Owens\cite{Owe87} has earlier
explained similar discrepancies in the photon production data of $pp$,
$\pi^+ p$, and $\pi^-p$ reactions at 19.4 GeV$<\sqrt{s}<63$ GeV as
arising from the parton intrinsic momentum \cite{Owe87}.  We shall provide
further support of Owens' observations by analytical and numerical
studies.

In this manuscript, we first examine photon production in
nucleon-nucleon collisions and extend our considerations to
nucleus-nucleus collisions.  In nucleon-nucleus and nucleus-nucleus
collisions, there are shadowing effects for photon production in
nuclei.  Previous experimental investigations of photon production
using $\pi^-$ beams on nuclei suggest that the effective shadowing in
the production of photons can be represented by a target mass
dependence of the form $A^\alpha$ with $\alpha \sim 1.0$ \cite{Zie97},
which can be explained theoretically \cite{Guo96}.  Because $\alpha$
is close to unity, the effective shadowing for photon production in
nuclei is quite weak.  We shall not include the shadowing effect on
photon production in nucleus-nucleus collisions.

\section{Photon Production in the Hard-Scattering Model}

In the relativistic hard-scattering process, the production of a
photon in a hadron-hadron reaction ($h_a+h_b \rightarrow \gamma+X$)
arises from the direct interaction of a parton of the hadron $h_a$
with a parton of the other hadron $h_b$.  The basic processes are the
Compton process $g+q ~({\rm or~} \bar{q}) \rightarrow \gamma+q ~({\rm
or~} \bar{q})$, and the annihilation processes $q \bar{q} \rightarrow
\gamma g$ \cite{Hwa85,McL85,Kaj86,Won94}.

One represents the probability for finding a parton $a$ with momentum
$a$=$(x_a,{\bbox {a}}_T)$ in the hadron $h_a$ by the parton distribution
function $G_{a/h_a}(x_a,{\bbox {a}}_T )$, which depends also on the
momentum transfer $Q^2$.  The $x_a$ and $Q^2$ dependences of the
parton distribution function $G_{a/h_a}$ for various partons have been obtained
by fitting large sets of experimental data in different
hard-scattering processes \cite{Duk84,CTEQ93,CTEQ95,Mar96}.

In leading-order perturbative QCD, the cross section for
$h_a+h_b\rightarrow \gamma+X$ is a convolution of the parton momentum
distributions and their elementary collision cross sections
$E_{\gamma} {d^3 \sigma}(a ~ b \rightarrow \gamma X')/{d p_{\gamma}^3}
$,
\begin{eqnarray}
\label{eq:hs}
E_{\gamma}& &{d^3 \sigma (h_a h_b \rightarrow \gamma X) \over {d
p_{\gamma}^3}}
=
\sum_{a b} \int  d x_a d \bbox{a}_T d x_b d \bbox{b}_T
\nonumber\\
& &\times G_{a/h_a} (x_a, \bbox{a}_T) G_{b/h_b} (x_b, \bbox{b}_T)
E_{\gamma} \frac{d^3 \sigma (a  b \rightarrow \gamma  X')}
{d p_{\gamma}^3}  .
\end{eqnarray}
The parton invariant cross section is related to $d\sigma/dt$ by
\begin{eqnarray}
\label{eq:dsdt}
&E_{\gamma}& \!\!\!\frac{d^3 \sigma (a  b \rightarrow \gamma  X')}
{d p_{\gamma}^3} 
\!= \!{d\sigma(a  b \rightarrow \gamma  X') \over \pi dt} 
\delta( \hat s + \hat t + \hat u - \!\!\sum_{i=1}^4 m_i^2 ) \nonumber\\
& &~~~~~~~~~~ \times{ \!\hat s \sqrt { \{ \hat s - (m_1+m_2)^2 \} 
\{ \hat s - (m_1-m_2)^2 \}}
\over (\hat s + m_4^2-m_3^2) } \,,
\end{eqnarray}
where $m_1$, $m_2$ are the masses of the incident partons, and $m_3$,
and $m_4$ are the masses of the outgoing particles.  In the above
equation, we have used the Mandelstam variables
\begin{eqnarray}
\hat{s} &= (a + b)^2 , \nonumber \\
\hat{t} &= (a - p_{\gamma})^2 , \\
\hat{u} &= (b - p_{\gamma})^2 .\nonumber
\end{eqnarray}

For definiteness, we shall use the nucleon-nucleon center-of-mass
system to refer to the momenta of the photon and the partons.  After
averaging the flavors, the colors, and the isospins of the initial
states of $q$, $\bar q$, and $g$, and summing over final states, the
differential cross section $d \sigma( a + b \rightarrow \gamma
+X')/dt$ are
\begin{eqnarray}
\label{eq:bxs1}
\frac{d {\sigma}(\!g q  \rightarrow \gamma q \!)   }
{d \hat{t}}
\!&=&\!{1\over 6}  \biggl (  \!\! \frac{e_q}{e}  \!\biggr )^2 
\!\frac{8\pi \alpha_s\alpha_e}{(\hat{s} - m^2)^2}
\biggl \{ \!\!\biggl ( \!\frac{m^2}{\hat{s}-m^2} +
\frac{m^2}{\hat{u}-m^2} \!\biggr )^2
\nonumber \\
+\biggl (\frac{m^2}{\hat{s}-m^2} &+& \frac{m^2}{\hat{u}-m^2} \biggr )
- \frac{1}{4} \biggl ( \frac{\hat{s}- m^2}{\hat{u}-m^2} +
\frac{\hat{u}- m^2}{\hat{s}-m^2} \biggr ) \!\!\biggr \}\,,\!\!\!\!
\end{eqnarray}
and
\begin{eqnarray}
\label{eq:bxs2}
\frac{d {\sigma} (\! q \bar{q}  \rightarrow \gamma g \! )}{d \hat{t}} 
\!&=& \!-{4\over 9}  \biggl( \!\!\frac{e_q}{e}\! \biggr )^2 \!\!
\! \frac{8\pi \alpha_s \alpha_e}{\hat{s}(\hat{s}-4m^2)}
\biggl \{ \!\!\biggl ( \frac{m^2}{\hat{t}-m^2} +
\frac{m^2}{\hat{u}-m^2}\! \biggr )^2
\nonumber \\ 
+ \biggl (\frac{m^2}{\hat{t}-m^2} &+& \frac{m^2}{\hat{u}-m^2} \biggr )
- \frac{1}{4} \biggl ( \frac{\hat{t}- m^2}{\hat{u}-m^2} +
\frac{\hat{u}- m^2}{\hat{t}-m^2} \biggr )  \biggr \}\,,\!\!\!\!
\end{eqnarray}
where $m$ is the quark mass, $e_q$ and $e$ are the quark charge and
the proton charge respectively, and $\alpha_{e}$ and $\alpha_s$ are
the electromagnetic fine-structure constant and the strong interaction
coupling constant, respectively.  
The strong coupling constant $\alpha_s$ is related to the 
 momentum transfer $Q^2$ by 
\begin{equation}
\alpha_s (Q^2) =
\frac{12 \pi}{(33 - 2 N_f){\rm ln} (Q^2/\Lambda^2)} ,
\end{equation}
where $N_f$ is the number of flavors, and $\Lambda$ is the QCD scale
parameter.

We shall assume a factorizable parton distribution function where the intrinsic
transverse momentum distribution can be factored out in the form
\begin{equation}
G_{a/h_a}(x_a, \bbox{a}_T) = F_{a/h_a}(x_a, Q^2) D_a(\bbox{a}_{T}).
\end{equation}
For convenience, we normalize the transverse momentum distribution
$D_a(\bbox{a}_{T})$ such that 
\begin{equation}
\label{eq:nor}
\int d{\bbox {a}}_T D_a(\bbox{a}_{T}) = 1.
\end{equation}
Then, the function $F_{a/h_a}(x_a, Q^2)$ is the usual parton
distribution function without assuming a  parton intrinsic transverse
momentum distribution, as given by \cite{Duk84,CTEQ93,CTEQ95,Mar96}.

\section{Effect of Parton Intrinsic Transverse Momentum}

Quarks, antiquarks, and gluons are asymptotically free at high
momenta, but their properties at low momenta are governed by the
confinement of these particles inside a hadron.  As a consequence, the
transverse momentum distribution of these particles is given by their
momentum wave functions.  The width of the momentum distribution is
related to the inverse of the size of the radius of confinement.
Thus, if one takes a confinement radius of 0.5 fm, the standard
deviation $\sigma$ of the parton intrinsic momentum should be of the
order of $\hbar /(0.5 {\rm ~fm})$=$0.4$ GeV/c, and the width of the
intrinsic transverse momentum about $\sqrt{2} \times 0.4 $GeV/c=0.65
GeV/c.

In the usual PQCD calculations, one neglects the parton intrinsic
transverse momentum by approximating the parton transverse momentum
distribution $D({\bbox{a}}_T)$ with a delta function
$D({\bbox{a}}_T)=\delta(\bbox{a}_T).$ Then, the integration over
$\bbox{a}_T$ and $\bbox{b}_T$ in Eq. (\ref{eq:hs}) can be trivially
carried out.  In such an approximate description, the colliding
partons have zero transverse momentum.  The produced photon acquires a
transverse momentum through the basic collision processes of $gq ({\rm
or~}\bar q) \rightarrow \gamma q ({\rm or~}\bar q)$ and $q\bar q
\rightarrow \gamma g$, with differential distributions as given by
Eqs.\ ({\ref{eq:bxs1}) and ({\ref{eq:bxs2}).

The transverse momentum of the photons produced from the collisions of
the constituents of the quark-gluon plasma expected in high-energy
heavy-ion collisions lies predominantly in the low $p_{\gamma T}$
region, extending to the region of intermediate $p_{\gamma T}$ of a
few GeV/c.  Hard-scattering processes in nucleon-nucleon collisions
also produce photons in this intermediate transverse momentum region.
It is necessary to determine the hard-scattering contributions to the
photon production in order to extract information on the quark-gluon
plasma photons.  The determination of the hard-scattering photons will
require the inclusion of the effects of the intrinsic parton
transverse momentum.

Previously, the effect of parton intrinsic transverse momentum has
been investigated by Owens \cite{Owe87} in hadron-nucleon collisions
at different energies.  In the range of 3 GeV/c $<p_{\gamma T} < 8$
GeV/c, Owens found by numerical calculations that the parton intrinsic
$p_T$ leads to an enhancement of the photon production cross section
and the enhancement increases as the incident hadron energy decreases
\cite{Owe87}.  The enhancement of photon production cross section due
to the parton intrinsic transverse momentum has also been reported in
recent numerical calculations for hadron-proton collisions at
$\sqrt{s}=$31.7 and 38.8 GeV \cite{Apa97}.  It is instructive to show
analytically in this section how the intrinsic transverse momentum
enhances the photon production cross section.  We shall make
simplifying assumptions in this section in order to carry out the
six-dimensional integration in Eq.\ ({\ref{eq:hs}) analytically.

We consider the production of a photon with $y_\gamma =0$ and
transverse momentum $p_{\gamma T}$.  We introduce the photon
light-cone variable, $x_\gamma=(E_\gamma+p_{\gamma z} )/\sqrt{s}$.
At $y_\gamma=0$, we have $p_{\gamma z}=0$ and $x_\gamma=p_{\gamma
T}/\sqrt{s}$. The photon momentum can be represented by
\begin{eqnarray}
\gamma=(E_\gamma,~ {\bbox p}_{\gamma T},~p_{\gamma z})= 
(x_\gamma {\sqrt{s}},~ {\bbox p}_{\gamma T},~ 0)\,.
\end{eqnarray}
The momenta of the target parton $a$ and projectile parton $b$ can be
written as (see Eq.\ (4.10) and (4.11) of \cite{Won94})
\begin{eqnarray}
a= \biggl (x_a {\sqrt{s}\over 2} + { {a}_T^2 \over 2 x_a \sqrt{s}},
~ \bbox{a}_T, ~-x_a {\sqrt{s} \over 2 } + { {a}_T^2 \over 2 x_a
\sqrt{s} } \biggr ),
\end{eqnarray}
\begin{eqnarray}
b= \biggl (x_b {\sqrt{s}\over 2} + { {b}_T^2 \over 2 x_b \sqrt{s}},
~\bbox{b}_T, ~x_b {\sqrt{s}\over 2} - {{ b}_T^2 \over 2 x_b
\sqrt{s}} \biggr ).
\end{eqnarray}
The Mandelstam variable $\hat s$, $\hat t$, and $\hat u$ are then
\begin{eqnarray}
\hat s=x_a x_b s + {{a}_T^2 {b}_T^2 \over x_a x_b s}
- 2 \bbox{a}_T \cdot \bbox{b}_T,
\end{eqnarray}
\begin{eqnarray}
\label{eq:hatu}
\hat u=-\biggl ( x_b + {{b}_T^2 \over  x_b {s}} \biggr ) x_\gamma s
+ 2 \bbox{b}_T \cdot \bbox{p}_{\gamma T},
\end{eqnarray}
\begin{eqnarray}
\label{eq:hatt}
\hat t=-\biggl ( x_a + { {a}_T^2 \over  x_a {s}} \biggr ) x_\gamma s
+ 2 \bbox{a}_T \cdot \bbox{p}_{\gamma T}\,.
\end{eqnarray}
We take the partons to be massless.   The momentum conservation condition,
${\hat s + \hat t + \hat u }=0$, leads to a relation of $x_a$ in terms
of $x_b$ given by
\begin{eqnarray}
\label{eq:xa1}
x_a&+&{ a_T^2 \over x_a s}
= x_\gamma + {1 \over  x_b + {(b_T^2 /  x_b s)}  - x_\gamma}   
\nonumber\\
&\times& \biggl [\!  x_\gamma^2 
\!-\! {2 (\bbox{a}_T\!  +  \!\bbox{b}_T)\cdot \bbox{p}_{\gamma T}\! 
- \!2\bbox{a}_T
\cdot \bbox{b}_T \over s} 
\!+\! {x _a b_T^2 \over x_b s}
\!+\! {x _b a_T^2 \over x_a s}\!  \biggr ] \,.
\nonumber
\end{eqnarray}
To simplify the last two terms in the square bracket, we note that
$x_a \sim x_b$ (see Eq.\ (\ref{eq:appr}) below).  The above equation
becomes
\begin{eqnarray}
\label{eq:xaxb}
 x_a+{ a_T^2 \over x_a s} 
= x_\gamma+ 
{ |\bbox{p}_{\gamma T} - \bbox{a}_T - \bbox{b}_T |^2 
\over  \{ x_b+ { (b_T^2 /  x_b s)}  - x_\gamma  \} s} \,,
\end{eqnarray}
which can be solved for $x_a$ to obtain $x_a={ x}_{a}(x_b)$.  We
assume a parton distribution function of the form $F(x)=A(1-x)^n$.
The integration over the light-cone variables $x_a$ and $x_b$ in Eq.\
({\ref{eq:hs}) becomes
\begin{eqnarray}
\label{eq:16}
\int &dx_a& dx_b  A_a A_b (1-x_a)^{n_a} (1-x_b)^{n_b} {\hat s \over \pi}
{d\sigma \over dt} {\delta(x_a-{ x}_{a}(x_b))  \over s(x_b - x _\gamma)} 
\nonumber\\
&=& A_aA_b\int dx_b e^{f(x_b)} {\hat s \over \pi s (x_b - x_\gamma)} {d\sigma
\over dt}\,,
\end{eqnarray}
where 
\begin{eqnarray}
f(x_b)= n_a \ln \bigl \{ 1-x_{a}(x_b) \bigr \} +  n_b \ln (1-x_b) .
\end{eqnarray}
The function $f(x_b)$ is an extremum of $x_b$ at 
\begin{eqnarray}
\label{eq:xb}
 x_{b}+{ b_T^2 \over x_{b} s} 
= x_\gamma+ \sqrt{ { n_a (1-x_{b}) \over n_b (1-x_{a}) }}
{|\bbox{p}_{\gamma T} - \bbox{a}_T - \bbox{b}_T | \over \sqrt{s} }\,.
\end{eqnarray}
The corresponding $x_{a}$ is given from
Eq.\ ({\ref{eq:xaxb}) by
\begin{eqnarray}
\label{eq:xa}
 x_{a}+{ a_T^2 \over x_{a} s} 
= x_\gamma+ \sqrt{ { n_b (1-x_{a}) \over n_a (1-x_{b}) }}
{ |\bbox{p}_{\gamma T} - \bbox{a}_T - \bbox{b}_T | \over \sqrt{s}} \,.
\end{eqnarray}
The integration over $x_a$ and $x_b$ in Eq.\ (\ref{eq:16}) becomes
\begin{eqnarray}
\label{eq:intx}
&&\int dx_a dx_b  A_a A_b (1-x_a)^{n_a} (1-x_b)^{n_b} {\hat s \over \pi}
{d\sigma \over dt} \delta(\hat s+ \hat t +\hat u)\nonumber\\
&=&{\sqrt{2 \pi \over B }}
{A_a A_b \hat s \over \pi s (x_{bT}-x_\gamma)} 
(1-x_{aT})^{n_a} (1-x_{bT})^{n_b}
{d\sigma \over dt}, 
\end{eqnarray}
where $x_{bT}$ and $x_{aT}$ are the solution of $x_b$ and $x_a$ from
solving Eqs.\ ({\ref{eq:xb}) and ({\ref{eq:xa}) respectively, and $B$
is
\begin{eqnarray}
B&=&-{\partial^2 f  \over \partial x_b^2} \biggr |_{x_b=x_{bT}}
\nonumber\\
&=&{n_b\over n_a}{2 \sqrt{n_a n_b}(1-2x_\gamma)+x_\gamma(n_a+n_b) \over
x_\gamma (1-2 x_\gamma)^2 }.
\end{eqnarray}
Consider the case with $p_{\gamma T} \! >> \! \sigma$, where $\sigma$
is the standard deviation of the parton $p_T$ distribution.  We can
expand in powers of the intrinsic transverse momenta to get
\begin{eqnarray}
\label{eq:pt}
|\bbox{p}_{\gamma T} \!-\! \bbox{a}_T \!-\! \bbox{b}_T|
\sim p_{\gamma T}\biggl (\! 1 - {\bbox{p}_{\gamma T}\cdot( \bbox{a}_T
+\bbox{b}_T) \over p_{\gamma T}^2  }\!  \biggr )\!+\!... \,.
\end{eqnarray}
Neglecting terms of second power in $\bbox{a}_T$ and $\bbox{b}_T$ and
using Eqs.\ (\ref{eq:xb}), (\ref{eq:xa}), and (\ref{eq:pt}), we can
separate out $x_{aT}$ and $x_{bT}$ into
\begin{eqnarray}
\label{eq:xat}
x_{i T}=x_{ i 0}+\Delta_i  ~~~~~~~~(i=a,b),
\end{eqnarray}
where 
\begin{eqnarray}
x_{a0}=\biggl (1+\sqrt{{ n_b (1-x_{a0}) \over n_a (1-x_{b0})}} 
\biggr )x_\gamma,
\end{eqnarray}
\begin{eqnarray}
\Delta_a 
= -\sqrt { {n_b(1-x_{a0})\over n_a(1-x_{b0})} } 
{ \bbox{p}_{\gamma T}\cdot( \bbox{a}_T + \bbox{b}_T ) \over 
p_{\gamma T} \sqrt{s}},
\end{eqnarray}
\begin{eqnarray}
x_{b0}=\biggl (1+\sqrt{{ n_a (1-x_{b0}) \over n_b (1-x_{a0})}}\biggr )
x_\gamma,
\end{eqnarray}
and 
\begin{eqnarray}
\Delta_b
=- \sqrt { {n_a(1-x_{b0}) \over n_b(1-x_{a0})}} 
{ \bbox{p}_{\gamma T}\cdot( \bbox{a}_T + \bbox{b}_T ) \over 
p_{\gamma T} \sqrt{s}}.
\end{eqnarray}
For the Compton process ($gq \rightarrow \gamma q $), $n_a \sim 6$ and
$n_b \sim 4$ \cite{Duk84}, and for the annihilation process ($q\bar q
\rightarrow \gamma g$), $n_a=n_b \sim 4$.  Therefore we have
approximately
\begin{eqnarray}
\label{eq:appr}
x_{a0} \sim x_{b0} \sim 2 x_\gamma.
\end{eqnarray}

We are now ready to carry out the integration over the intrinsic
transverse momenta $\bbox{a}_T$ and $\bbox{b}_T$ in Eq.\
(\ref{eq:hs}).  The dependence of $(1-x_{aT})^{n_a}$ on the intrinsic
momentum $\bbox{a}_T$ can be separated out to be
\begin{eqnarray}
(1&-&x_{aT})^{n_a} ~ \sim ~(1-x_{a0})^{n_a}
\exp\{-{n_a \Delta_a \over 1 - x_{a0}}\}
\nonumber\\
& & \!\!\!\!\!\!\! \sim  (1-x_{a0})^{n_a} \exp \biggl \{ {\sqrt{n_a n_b} 
{ \bbox{p}_{\gamma T}\cdot( \bbox{a}_T + \bbox{b}_T ) \over 
(1 - x_{a0})  p_{\gamma T} \sqrt{s}} }.
   \biggr \}
\end{eqnarray}
The factor $(1-x_{bT})^{n_b}$ can be similarly separated.
The integral over the transverse momentum in Eq.\ (\ref{eq:hs}) becomes
\begin{eqnarray}
\label{eq:ptint}
\!\!\! & & \!\!\int \!\! d \bbox{a}_T d \bbox{b}_T
D_a (\bbox{a}_T) D_b (\bbox{b}_T)
(1-x_{aT})^{n_a} (1-x_{bT})^{n_b}
{d\sigma \over dt} 
\nonumber\\
&=&(1-2x_\gamma)^{n_a+n_b}
\int d{\bbox a}_T d{\bbox b}_T D_a ({\bbox a}_T)D_b ({\bbox b}_T) 
\nonumber\\
& &~~~~~~~~~\times
\exp \biggl \{ 2 {\sqrt{n_a n_b} 
{ \bbox{p}_{\gamma T}\cdot( \bbox{a}_T + \bbox{b}_T ) \over 
(1 - 2x_\gamma)  p_{\gamma T} \sqrt{s}}} \biggr \} {d\sigma \over dt},
\end{eqnarray}
where we have used Eq.\ ({\ref{eq:appr}).  We assume a Gausssian
parton transverse momomentum distribution, $D(\bbox{p}_T)=
\exp\{-p_T^2/2\sigma^2\}/2\pi \sigma^2$ with $<\!{p}_T^2\!>=2\sigma^2$,
and we obtain
\begin{eqnarray}
\label{eq:inta}
& &\!\!\!\!\!\!\!\!\!\int \!\!\!d{\bbox a}_T d{\bbox b}_T 
D_a (\!{\bbox a}_T \! )D_b ({\bbox b}_T \! ) 
\exp \biggl \{\! 2 {\sqrt{n_a n_b} 
{ \bbox{p}_{\gamma T} \!\cdot \!( \bbox{a}_T \!+\! \bbox{b}_T ) \over 
(\sqrt{s} \!-\! 2p_{\gamma T})  p_{\gamma T} }}\! \biggr \}\! 
{d\sigma \over dt} \nonumber \\ 
\!\!&=&\!  \exp \biggl \{ \!\!{4\sigma^2 n_a n_b \over (\sqrt{s}\!-\!2
p_{\gamma T})^2 } \!\!\biggr \} 
\!\!\int \!\!d{\bbox a}_T d{\bbox b}_T D_a( {\bbox a}_T 
\!-\! \bbox{\lambda}) 
D_b( {\bbox b}_T \!-\! \bbox{\lambda})\! {d\sigma \over dt}\!\! \,,
\nonumber\\
{}
\end{eqnarray}
where 
\begin{eqnarray}
{\bbox \lambda} = 
{2 \sigma^2 \sqrt{n_a n_b}~ \bbox{p}_{\gamma T} 
\over ( \sqrt{s}-2p_{\gamma T})p_{\gamma T} }.
\end{eqnarray}
The first factor on the right-hand side of Eq.\ ({\ref{eq:inta}),
\begin{eqnarray}
\label{eq:kappa1}
\kappa_1 = \exp \biggl \{ {4\sigma^2 n_a n_b \over (\sqrt{s}-2
p_{\gamma T})^2 } \biggr \},
\end{eqnarray}
is the enhancement factor due to the parton intrinsic transverse
momentum and the variation of the parton distribution function.  We
can understand the origin of this enhancement in the following way.
The integration over $x_a$ and $x_b$ leads to a cross section
proportional to $(1-x_{aT})^{n_a} (1-x_{bT})^{n_b}$ (Eq.\
(\ref{eq:intx})).  If there were no parton intrinsic transverse
momentum, $x_{aT}$ would be given by $x_{a0}$.  The presence of the
intrinsic transverse momentum leads to $x_{aT}$ varying about $x_{a0}$
with the variation $\Delta_a$ depending on the intrinsic transverse
momentum (Eq.\ ({\ref{eq:xat})).  Because of the steep decrease of
$(1-x_a)^{n_a}(1-x_b)^{n_b}$ as a function of $x_a$ and $x_b$, the
averaging over the transverse momenta leads to the enhancement factor
$\kappa_1$.

The enhancement factor $\kappa_1$ arising from the parton distribution
function allows us to determine the scale of $p_{\gamma T}$ at which
the effect of the parton intrinsic $p_T$ will be important.  The
enhancement factor $\kappa_1$ is close to unity if
$2<\!\!p_T^2\!\!>\!n_a n_b\!\!  <<\!\!  (\sqrt{s}\!-\!2 p_{\gamma
T})^2$ and is much greater than unity if $2<\!p_T^2\!>\!n_a n_b$ is
large or comparable to $(\sqrt{s}-2 p_{\gamma T})^2$.  As $n_a n_b
\sim 24$ for the Compton process, the importance of the intrinsic
$p_T$ depends on whether $7 \sqrt{<\!p_T^2\!>}$ is small compared to
$\sqrt{s}-2 p_{\gamma T}$.  When $7 \sqrt{<\!p_T^2\!>}$ is large or
comparable to $(\sqrt{s}-2 p_{\gamma T})$, the enhancement factor
$\kappa_1$ becomes quite large.  This is the main origin for the large
enhancement factor for $p_{\gamma T}< 6$ GeV/c in hadron-proton
collisions at $\sqrt{s} \sim 20$ GeV.

For the cross section $d\sigma/dt$, the dominant contributions come
from terms proportional to $1/\hat t$ and $1/\hat u$.  It can be
approximated by
\begin{eqnarray}
\label{eq:s0}
{d \sigma \over dt}
\sim C \biggl ( { 1\over \hat t} + {1 \over \hat u} \biggr ),
\end{eqnarray}
From Eqs.\ (\ref{eq:hatu}), (\ref{eq:hatt}), and (\ref{eq:appr}), we
have
\begin{eqnarray}
\label{eq:s1}
\hat t \sim | \bbox{p}_{\gamma t} - \bbox {b}_T|^2,
\end{eqnarray}
and 
\begin{eqnarray}
\label{eq:s2}
\hat u \sim | \bbox{p}_{\gamma t} - \bbox {a}_T|^2.
\end{eqnarray}
The integration over the intrinsic transverse momentum distributions
can be carried out to yield
\begin{eqnarray}
\label{eq:intt}
\!\!\!\int& &\!\!\!d{\bbox a}_{{}_T}\! d{\bbox b}_{{}_T} 
D_a( {\bbox a}_{{}_T} \!\!-\!\bbox{\lambda}) 
D_b( {\bbox b}_{{}_T} \!\!-\!\bbox{\lambda})
C\biggl (\! {1 \over  ( \bbox p_{\gamma {{}_T}}\!\!\! - \!
\bbox{a}_{{}_T}\!)^2}  
\!+\! {1 \over  ( \bbox p_{\gamma {{}_T}} \!\!\!-\! 
\bbox{b}_{{}_T}\!)^2}\!\! \biggr )  
\nonumber\\
& &~\sim~ \biggl \{ {p_{\gamma {}_T}^2 \over 
( \bbox p_{\gamma {}_T} \!-\!
\bbox{\lambda})^2-2\sigma^2} \biggr \}     {2C \over p_{\gamma T}^2} \,,
\end{eqnarray}
where first factor on the right hand side
\begin{eqnarray}
\label{eq:kappa2}
\kappa_2=  {p_{\gamma T}^2 \over ( \bbox p_{\gamma T} -
\bbox{\lambda})^2-2\sigma^2},
\end{eqnarray}
is the enhancement factor due to the parton intrinsic transverse
momentum and the variation of $d\sigma/dt$.  This enhancement factor
$\kappa_2$ arising from $d\sigma/dt$ is close to unity if $p_{\gamma
T}\!>>\!  \sigma$.  From Eqs.\ (\ref{eq:s0}), (\ref{eq:s1}), and
(\ref{eq:s2}), the enhancement factor due to $d\sigma/dt$ becomes much
greater than unity when $p_{\gamma T}$ is of the order $\sqrt{<\!
p_T^2 \!>}$.

Putting all the results
together, we have
\begin{eqnarray}
\label{eq:final}
E_{\gamma} {d^3 \sigma (h_a h_b \rightarrow \gamma X) \over {d
p_{\gamma}^3}}~~~~~& & \nonumber \\
\sim \kappa 
{\sqrt{2 \pi \over B }}
{4 A_a A_b p_{\gamma T}  \over \pi \sqrt{s}} 
& &\biggl (1-{2 p_{\gamma T} \over \sqrt{s}} 
\biggr )^{n_a+n_b} { 2C \over p_{\gamma T}^2},
\end{eqnarray}
where $\kappa$ is given by 
\begin{eqnarray}
\label{eq:kappa}
\kappa = \kappa_1 \kappa_2=
\exp \biggl \{\! {4\sigma^2 n_a n_b \over (\sqrt{s}-2 p_{\gamma T})^2 }
\! \biggr \}
{p_{\gamma T}^2   \over p_{\gamma T}^2
(1-\lambda)^2 -2 \sigma^2 }.
\end{eqnarray}
The quantity $\kappa$ in Eqs.\ (\ref{eq:final}) and (\ref{eq:kappa})
is the ratio of the photon production cross section with the parton
intrinsic transverse momentum to the the cross section without the
parton intrinsic transverse momentum. From Eq.\ (\ref{eq:kappa}) we
infer that for $p_{\gamma T}\! >> \! \sqrt{<\!p_T^2\!>}$, which allows
the expansion in Eq.\ ({\ref{eq:pt}), $\kappa$ is greater than unity
and represents an enhancement due to the parton intrinsic transverse
momentum.  It is a function of both $\sqrt{s}$ and $p_{\gamma T}$.  It
decreases as $\sqrt{s}$ increases.  For the same $\sqrt{s}$, the first
factor increases with increasing $p_{\gamma T}$ while the second
factor increases with decreasing $p_{\gamma T}$.

If one uses $2 \sigma^2=0.9$ (GeV/c)$^2$ as given by Owens
\cite{Owe87}, then for $p_{\gamma T} = 6$ GeV/c at $\sqrt{s}=20$ GeV,
Eqs.\ (\ref{eq:kappa1}) and (\ref{eq:kappa2}) give $\kappa_1=2.2$ and
$\kappa_2=1.27$, and a combined enhancement factor $\kappa=2.8$, in
rough agreement with the enhancement factor of about $3.8$ obtained from
numerical calculations in Figure 3.  There is a substantial
enhancement of the photon production cross section at $\sqrt{s}\sim
20$ GeV due to the parton intrinsic transverse momentum.  For
$p_{\gamma T}=6$ GeV/c at $\sqrt{s}=63$ GeV, $\kappa$ decreases to
1.07, in agreement with the numerical results in Figure 4.  This value
of $\kappa$ is only slightly greater than unity, indicating that the
parton intrinsic $p_T$ has only small influence for $p_{\gamma T} \sim
6$ GeV/c at $\sqrt{s}> 63$ GeV.

The enhancement factor $\kappa$ in Eq.\ (\ref{eq:kappa}) shows that
one should expect a large enhancement due to the intrinsic transverse
momentum for the production of photons with transverse momentum
$p_{\gamma T}$ close to $\sqrt{s}/2$ such that $\sqrt{s}-2 p_{\gamma
T}$ is not much greater than $\sqrt{<\!p_T^2\!>}$.  In this case,
$x_{a0}\sim 2p_{\gamma T}/\sqrt{s}$ is close to unity and the
magnitude of the cross section is however quite small.

The above analytical integration of the six-dimensional integral of
Eq.\ ({\ref{eq:hs}) serves the purpose of illustrating the important
features of the enhancement due to the parton intrinsic transverse
momentum.  In order to compare with experimental data quantitatively,
we need to carry out numerical integrations of Eq.\ ({\ref{eq:hs}) in
their full complexities without simplifying assumptions in the
following sections.

\section{Transverse Momentum Distribution of Partons}

The momentum distribution of partons for small $p_T$ depends on
confinement, while the momentum distribution at high $p_T$ depends on
the constituent counting rule which is a power law in nature
\cite{Bla74,Siv76,Won94}.  They have different functional forms.  To
carry out our investigation of the parton intrinsic momentum in
numerical calculations, we join the two different functional forms at
$p_{TM}$ and assume the parton $p_T$ distribution function of the form
\begin{eqnarray} 
D (\bbox{p}_T) = C &\biggl [& \frac{1}{1 + e^{(p_T - p_{TM})/
\Delta}} D_1({p}_T) \nonumber\\
&+& \frac{1}{1 + e^{(p_{TM}- p_{T})/\Delta}} D_2({p}_T)\biggr ] ,
\end{eqnarray}
where $C\sim 1$ will be determined by the normalization condition
Eq. (\ref{eq:nor}).

Following Owens \cite{Owe87}, we use the Gaussian distribution
$D_1(\bbox{p}_T)$ for the distribution at low $p_T$ as given by
\begin{equation}
D_1({p}_T) = \frac{1}{\pi \langle p_T^{~2} \rangle} e^{- {p_T^2}/
{<\! p_T^{2} \!>}} .
\end{equation}
For large $p_T$, we take the transverse momentum distribution
$D_2({p}_T)$ to be given by a power-law distribution as in the
spectator counting rule of Blankenbecler and Brodsky \cite{Bla74}
\begin{equation}
D_2({p}_T) =         \frac{a}  {({p}_T^{~2} + M^2)^n},
\end{equation}
where $a=D_1( p_{TM}){({p}_{TM}^{~2} + M^2)^n}$.  The results of our
calculation for photon production are rather insensitive to this
transverse distribution $D_2$.  Using information from the work of
Sivers, Blankenbecler, and Brodsky \cite{Siv76}, we choose $n=4$ and
$M=1$ GeV to describe the transverse momentum distribution $D_2$.  For
numerical purposes, we set the matching transverse momentum to be
$p_{TM}=2\sqrt{<\! p_T^{2} \!>}$ , and the matching width
$\Delta=0.25$ GeV/c.

\section{Next-to-Leading-Order Perturbative QCD}

It is well known that higher-order perturbative QCD corrections are
important in the evaluation of the cross sections for many
hard-scattering processes (see for example, \cite{Fie89,CTEQ93}).  For
photon production, contributions up to the second order in $\alpha_s$
have been considered by Aurenche, Baier, Douiri, Fontannaz, and
Schiff, and by Baer, Ohnemus, and Owens \cite{Aur87,Aur88,Bae90}.
These calculations involve the evaluation of contributions from all
next-to-leading-order Feynman diagrams.  In these calculations
\cite{Aur87,Aur88,Bae90}, the intrinsic transverse momentum of the
partons has been neglected by using a delta function distribution
$\delta(\bbox{p}_T)$ so that integrations over the transverse momentum
of the partons can be trivially carried out.  From these leading-order
plus next-to-leading-order ($LO\!+\!NLO$) calculations, the effects of
the higher-order terms can be represented by the $K$-factor defined by
\begin{equation}
\label{eq:kfac}
K(p_{\gamma})
= \biggl \{ E_{\gamma} \frac{d^3 \sigma}{d p_{\gamma}^3}(LO\!+\!NLO) \biggl / 
E_{\gamma} \frac{d^3 \sigma}{d p_{\gamma}^3}(LO) \biggr \},
\end{equation}
where $E_{\gamma} (d^3 \sigma/{d p_{\gamma}^3})(LO\!+\!NLO)$ is the photon
obtained with the inclusion of both the lowest-order and the
next-to-leading-order terms, and $E_{\gamma} (d^3 \sigma/{d
p_{\gamma}^3})(LO)$ is the photon cross section obtained in the
lowest-order calculation.

Numerically, the $K$-factor is about 2 to 3.5 and is a slowly varying
function of the photon transverse momentum (see Figs. 1 and 2).  It
represents an effective modification of the strength of the coupling
constant at the vertex, as in a vertex correction, due to the
initial-state interaction between partons
\cite{Fie89,Cha94,Won96,Won97}.  On the other hand, the vertex
correction depends on the relative momentum between the colliding
partons \cite{Cha94,Won96,Won97}, which in turn depends mainly on the
longitudinal momenta; the vertex correction is insensitive to the
parton transverse momenta.  Because of this property, it is reasonable
to consider the $K$-factor to be approximately independent of the
parton transverse momentum; that is
\begin{eqnarray}
\biggl \{  &E_{\gamma}&\!\! \frac{d^3 \sigma}{d p_{\gamma}^3}
(LO \!+\! NLO \!+\! PT)
\biggl / E_{\gamma} \frac{d^3 \sigma}{d p_{\gamma}^3}(LO \!+\! NLO ) 
   \biggr \}  \nonumber\\ 
&\sim& 
\biggl \{E_{\gamma} \frac{d^3 \sigma}{d p_{\gamma}^3}(LO\!+\!NLO) \biggl /
E_{\gamma} \frac{d^3 \sigma}{d p_{\gamma}^3}(LO) \biggr \},
\end{eqnarray}
where the abbreviation ``$PT$'' stands for calculations where the
parton transverse momentum distribution is taken into account.  Upon
using the above approximation to include the effects of the
next-to-leading-order corrections and the parton intrinsic transverse
momentum distribution, the photon production cross section is given by
\begin{eqnarray}
\label{eq:app}
E_{\gamma} \frac{d^3 \sigma}{d p_{\gamma}^3}(LO\!+\!NLO\!+\!PT)
\!\sim\!
K(p_{\gamma}\!)E_{\gamma} \frac{d^3 \sigma}{d p_{\gamma}^3}(LO\!+\!PT) .
\end{eqnarray}

\epsfxsize=300pt
\includegraphics{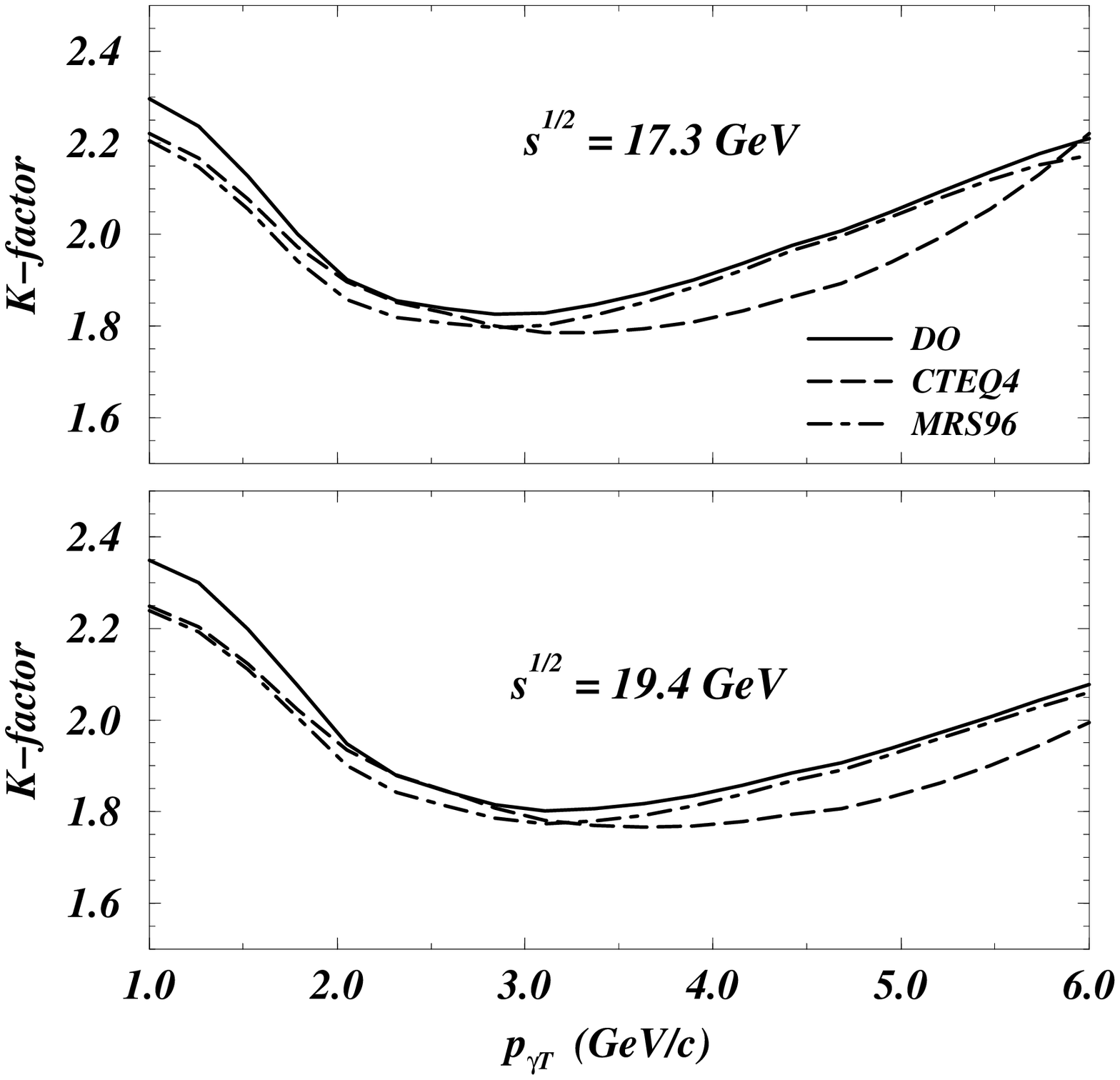}
\null\vskip 7.9cm
\begin{minipage}[t]{8cm}
\noindent {\bf Fig.\ 1}.  {The $K$-factor as a function of the photon
transverse momentum $p_{\gamma T}$ at $y_\gamma=0$.  ($a$) for $pp$
collisions at $\sqrt{s}=17.3$ GeV and ($b$) for $\sqrt{s}=19.4$ GeV.
}
\end{minipage}
\vskip 4truemm
\noindent 

\section{Photon Production in Nucleon-Nucleon Collisions} 

From Eq.\ ({\ref{eq:app}), we can take into account the effects of
intrinsic transverse momentum and next-to-leading-order contributions
by carrying out the calculation in two steps.  We perform the $LO\!+\!NLO$
calculations, without the intrinsic parton transverse momentum
distribution, using the numerical code from Aurenche $et~al.$
\cite{Aur87,Aur88}.  The ratio of the $LO\!+\!NLO$ calculation to the
leading-order ($LO$) calculation then gives the $K$-factor as a
function of the photon momentum, as defined by Eq.\ (\ref{eq:kfac}).
We then carry out separately a numerical calculation of the leading
order with the intrinsic transverse momentum distribution.  Equation
({\ref{eq:app}) is then used to give the cross section when both
effects are included.

The photon cross section depends on the functional dependence of $Q^2$
in terms of physical quantities such as $p_{\gamma T}$ of the photon.
Owens has studied different choices of the functional form, from
$Q^2=p_{\gamma T}^2$ to $Q^2=p_{\gamma T}^2/4$, and found that
$Q^2=p_{\gamma T}^2/2$ gives the best description of the experimental
data for production of photons with large transverse momenta
\cite{Owe87}.  We shall accordingly use this functional dependence for
our investigation.

We first carry out the $LO$ and the $LO\!+\!NLO$ calculations using three
different parton distribution functions: (1) the Duke and Owens parton
distribution function (set 1) \cite{Duk84}, (2) the CTEQ4 parton
distribution function (set 3) \cite{CTEQ93,CTEQ95}, and (3) the MRS96
parton distribution function (set 1) \cite{Mar96}.  The ratio of the
$LO\!+\!NLO$ result to the $LO$ result gives the $K$-factors which are
shown in Figs.\ 1 and 2, and the the results for $LO$ and $LO\!+\!NLO$
cross sections are shown in subsequent figures.  The $K$-factor in
Figs.\ 1 and 2 as a function of the photon transverse momentum
$p_{\gamma T}$ for $y_\gamma=0$ indicate that it has a value of
approximately 2 for 1 GeV/c $< p_{\gamma T} < 6$ GeV/c in
nucleon-nucleon collisions at $\sqrt{s}=$ 17.3 and 19.4 GeV.  It
behaves slightly differently for $\sqrt{s}=63 $ GeV, where it drops
from about 3 to 1.4, as $p_{\gamma T}$ decreases from 2 GeV/c to 6
GeV/c.  It becomes nearly a constant for $p_{\gamma T}>$ 6 GeV/c.

\epsfxsize=300pt
\includegraphics{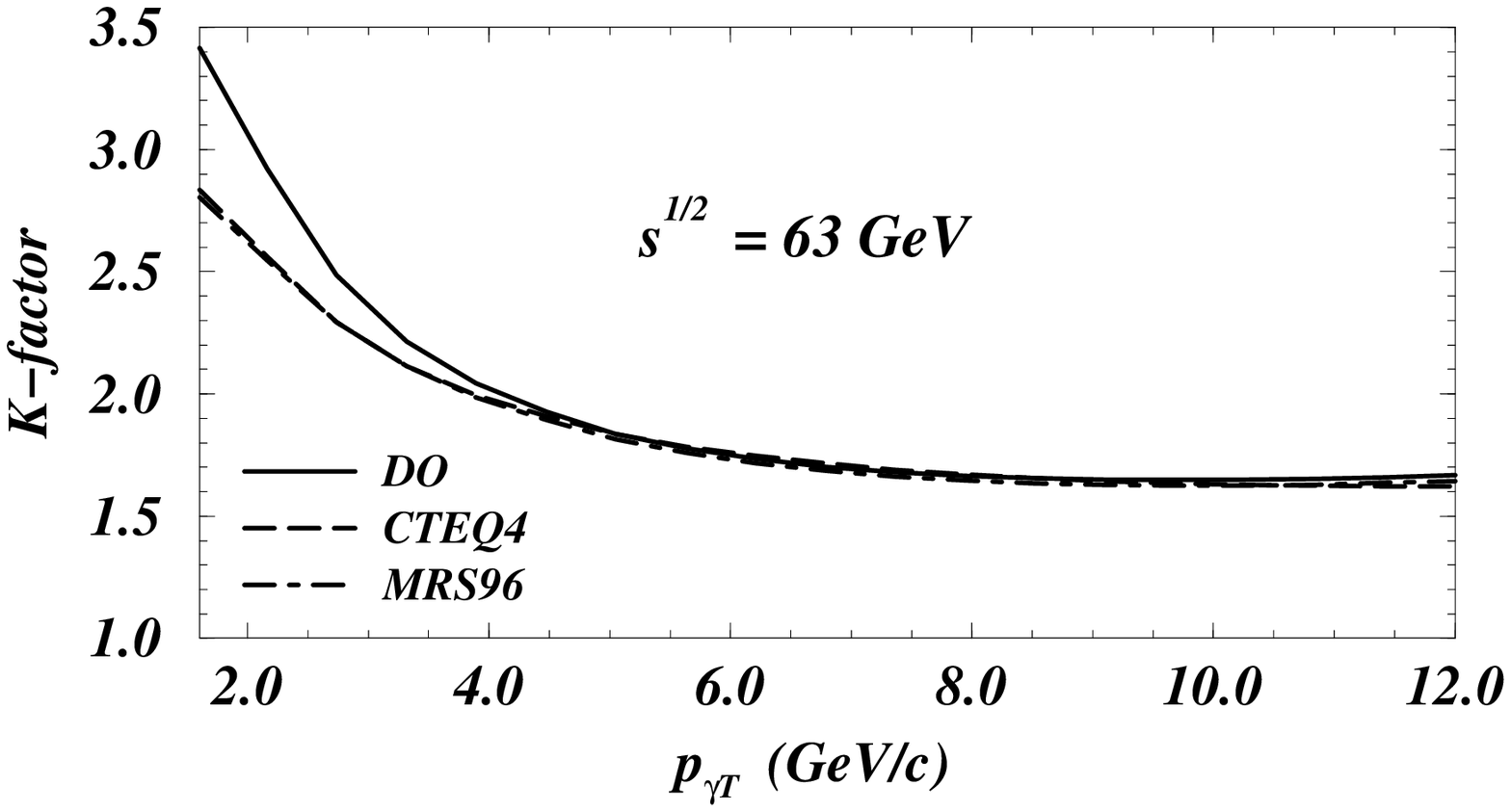}
\vskip 5.2cm
\begin{minipage}[t]{8cm}
\noindent {\bf Fig.\ 2}.  { The $K$-factor as a function of the photon
transverse momentum $p_{\gamma T}$ at $y_\gamma=0$ for $pp$ collisions
at $\sqrt{s}=63$ GeV.  }
\end{minipage}
\vskip 4truemm
\noindent 

We next carry out a lowest-order calculation with the inclusion of the
parton transverse momentum distribution of the partons.  Previously,
Owens noted that the transverse momentum distribution of dilepton
pairs gives a direct measure of the parton intrinsic transverse
momentum \cite{Owe87}.  Data obtained by Kaplan $et~al.$\cite{Kap78}
for $pp$ collisions at $\sqrt{s}=27.4$ GeV indicate
$<p_T^2(\mu^+\mu^-)>=1.9$ (GeV/c)$^2$ which corresponds to a parton
transverse momentum $<\! p_T^2 \!>=0.95$ (GeV/c)$^2$ in the lowest
order approximation.  The value of $<\!  p_T^2 \!>$ decreases if
higher-order terms are included \cite{Fie89}.  The convolution method
of \cite{Fie89} is however a model-dependent prescription for
incorporating the effects of the intrinsic transverse momentum.
Nevertheless, it indicates that the amount of intrinsic transverse
momentum needed to describe the data is influenced by the amount of
QCD dynamics included in the analysis.  The analysis of Owens using
the lowest-order QCD with a constant $K$-factor indicates that $<\!
p_T^2 \!>=0.9$ (GeV/c)$^2$ gives a good agreement with photon
production data in $pp$, $\pi^+p$ and $\pi^- p$ collisions at 19.4
GeV$<\sqrt{s}<$62 GeV \cite{Owe87}.  Recent analysis \cite{Apa97}
suggests even a greater value of $<\!  p_T^2 \!>$ of about 2
(GeV/c)$^2$. More future work is needed to clarify the situation on
the the magnitude of the parton intrinsic transverse momentum.

\null\protect\vspace*{1cm}
\epsfxsize=300pt
\includegraphics{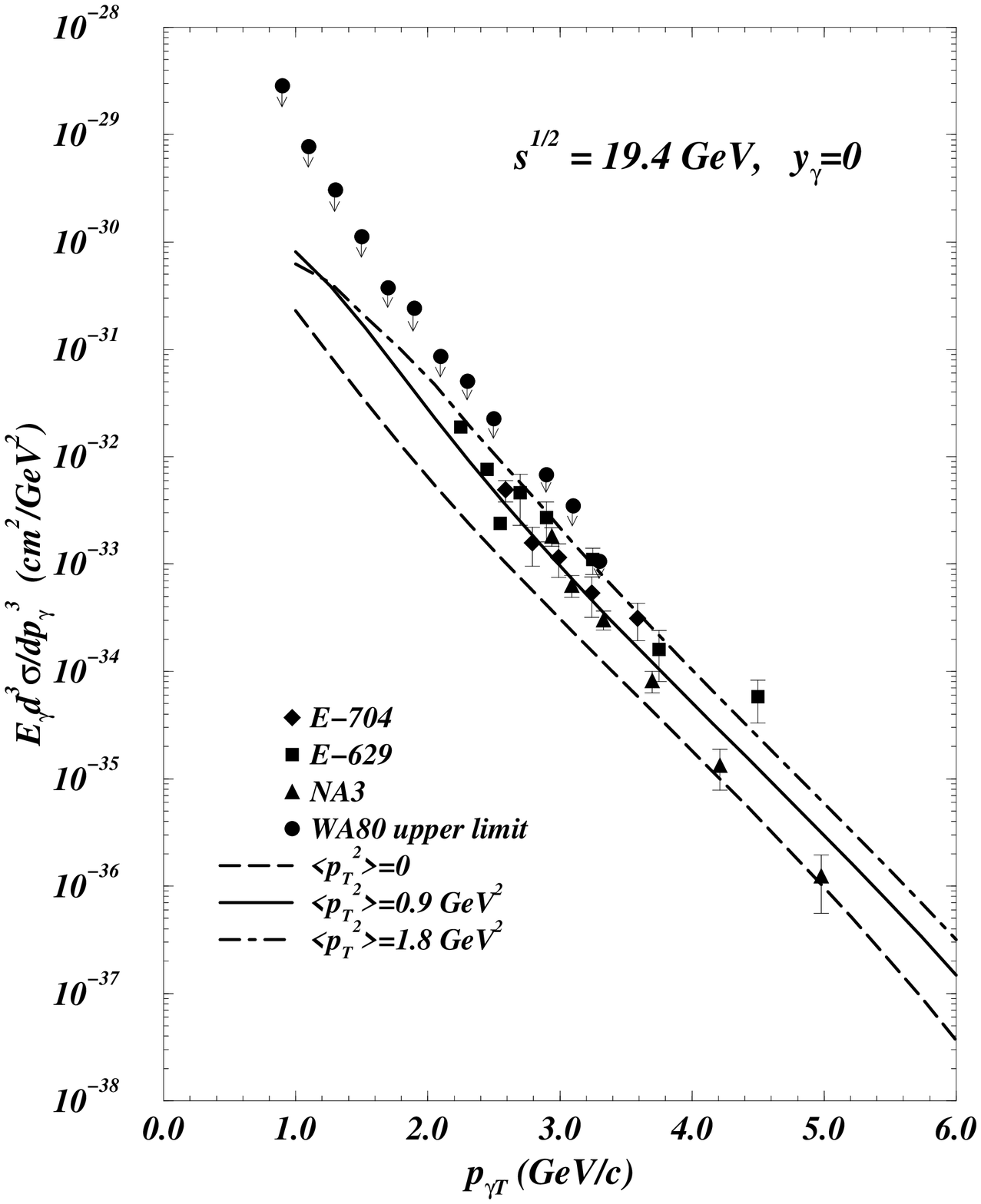}
\vskip 8.5cm
\begin{minipage}[t]{8cm}
\noindent {\bf Fig.\ 3}.  {The photon invariant cross section for
nucleon-nucleon collisions at $\sqrt{s}=19.4$ GeV and $y_\gamma=0$
calculated with different $<\! p_T^2 \!>$ values of parton intrinsic
momentum distribution and the Duke and Owens parton distribution
function (set 1).  The $NLO$ corrections have been included.}
\end{minipage}
\vskip 4truemm
\noindent

To see the dependence of the photon transverse momentum distribution
on the magnitude of $<\! p_T^2 \!>$ of the parton intrinsic transverse
momentum, we show in Fig.\ 3 the results for $pp$ collisions at
$\sqrt{s}=19.4$ GeV, using the Duke and Owens parton distribution
function (set 1).  The cross sections for $<\! p_T^2 \!>=0 $ are less
than those for $<\! p_T^2 \!>=0.9$ (GeV/c)$^2$ by about a factor of 3
to 4, which are less than those for $<\! p_T^2 \!>$=1.8 (GeV/c)$^2$ by
about a factor of 2 (except at $p_{\gamma T} \simeq$ 1 GeV/c).  There
is a substantial enhancement of the cross section due to the parton
intrinsic transverse momenta.  The value of $<\! p_T^2 \!>=0.9$
(GeV/c)$^2$ provides a good description of the experimental data.  The
enhancement factor for $<\! p_T^2 \!>=0.9$ (GeV/c)$^2$ is relatively
constant for $\sqrt{s}\sim 20$ GeV.

\null\protect\vspace*{0.2cm}
\epsfxsize=300pt
\includegraphics{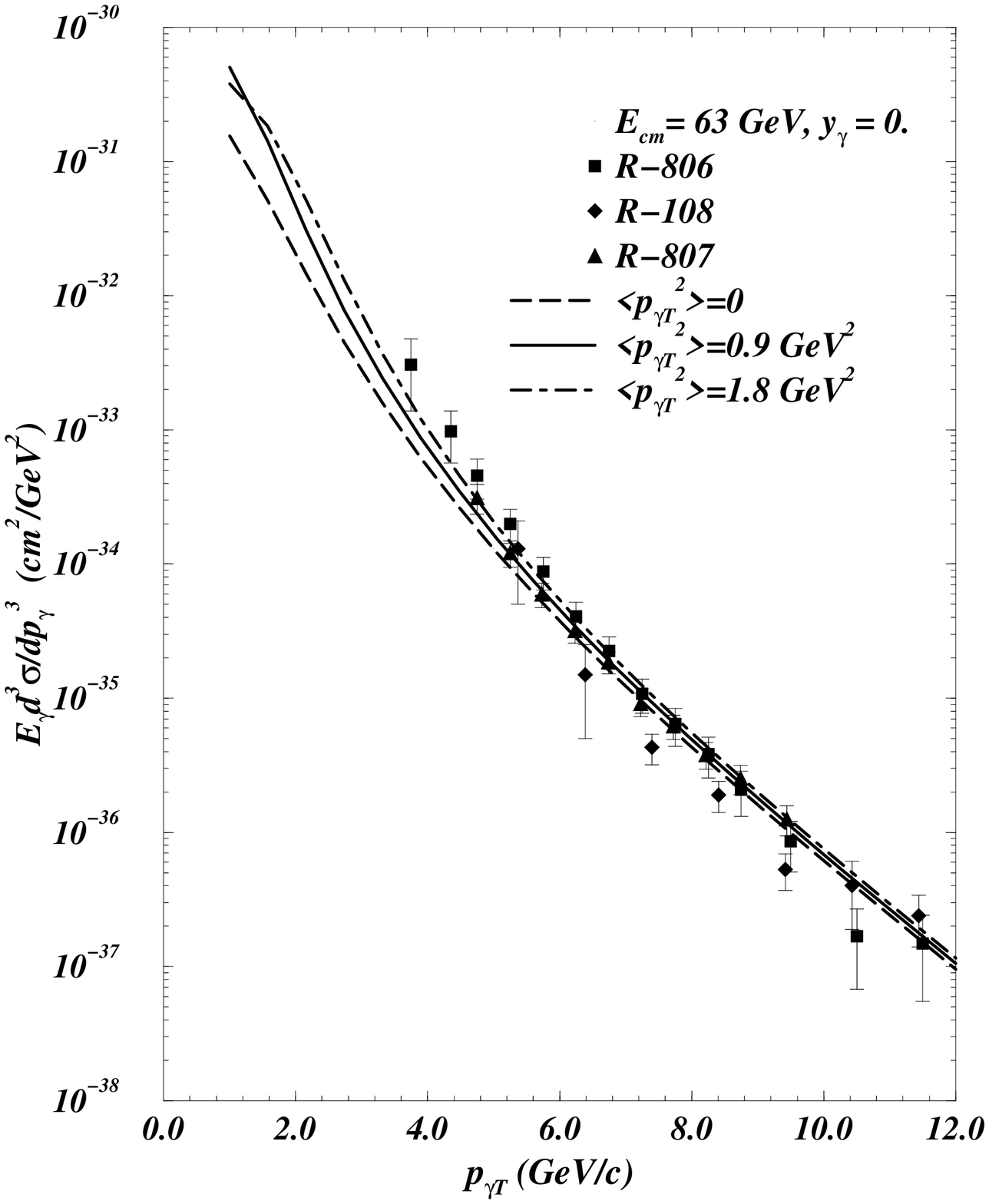}
\vskip 9.1cm
\begin{minipage}[t]{8cm}
\noindent {\bf Fig.\ 4}. {The photon invariant cross section for
nucleon-nucleon collisions at $\sqrt{s}=63$ GeV and $y_\gamma=0$
calculated with different $<\! p_T^2 \!>$ values of parton intrinsic
momentum distribution and the Duke and Owens parton distribution
function (set 1).  The $NLO$ corrections have been included.}
\end{minipage}
\vskip 7truemm
\noindent 

The enhancement of the photon production cross section in the region
of $p_{\gamma T}$ under consideration poses an interesting puzzle.  If
the running coupling constant could be taken as a fixed constant, the
total cross section obtained by integrating the photon momentum should
be independent of the width (or $<\! p_T^2 \!>$) of the parton
intrinsic transverse momentum distribution.  An enhancement of the
cross sections in all regions of $p_{\gamma T}$ appears to violate
this normalization of the total photon production cross section.  The
puzzle is resolved by noting that the cross section is not enhanced in
all regions of $p_{\gamma T}$.  The photon production cross section is
enhanced in the region $p_{\gamma T} \! >> \!  \sqrt{<\! p_T^2 \!>}$,
and it is suppressed for $p_{\gamma T} << \sqrt{<\! p_T^2 \!>}$
because partons acquire energies through their intrinsic $p_T$ and the
probability for producing very low energy photons therefore decreases
when partons have intrinsic $p_T$.  A hint of the suppression for low
$p_{\gamma T}$ can be found in Figs. 3 and 4 where for $p_{\gamma T}
<1.2$ GeV/c the cross section for the case $<\!  p_T^2 \!>=1.8$
(GeV/c)$^2$ is lower than that for $<\!  p_T^2 \!>=0.9$ (GeV/c)$^2$.
While we expect a suppression for this region of small $p_{\gamma T}\!
<< \!  \sqrt{<\!p_T^2\!>}$, it is a different matter whether
perturbative QCD can remain valid for this region where the associated
value of $Q^2$ becomes very small.  We shall not consider this region
of small $p_{\gamma T}$.

\null\protect\vspace*{1.0cm}
\epsfxsize=300pt
\includegraphics{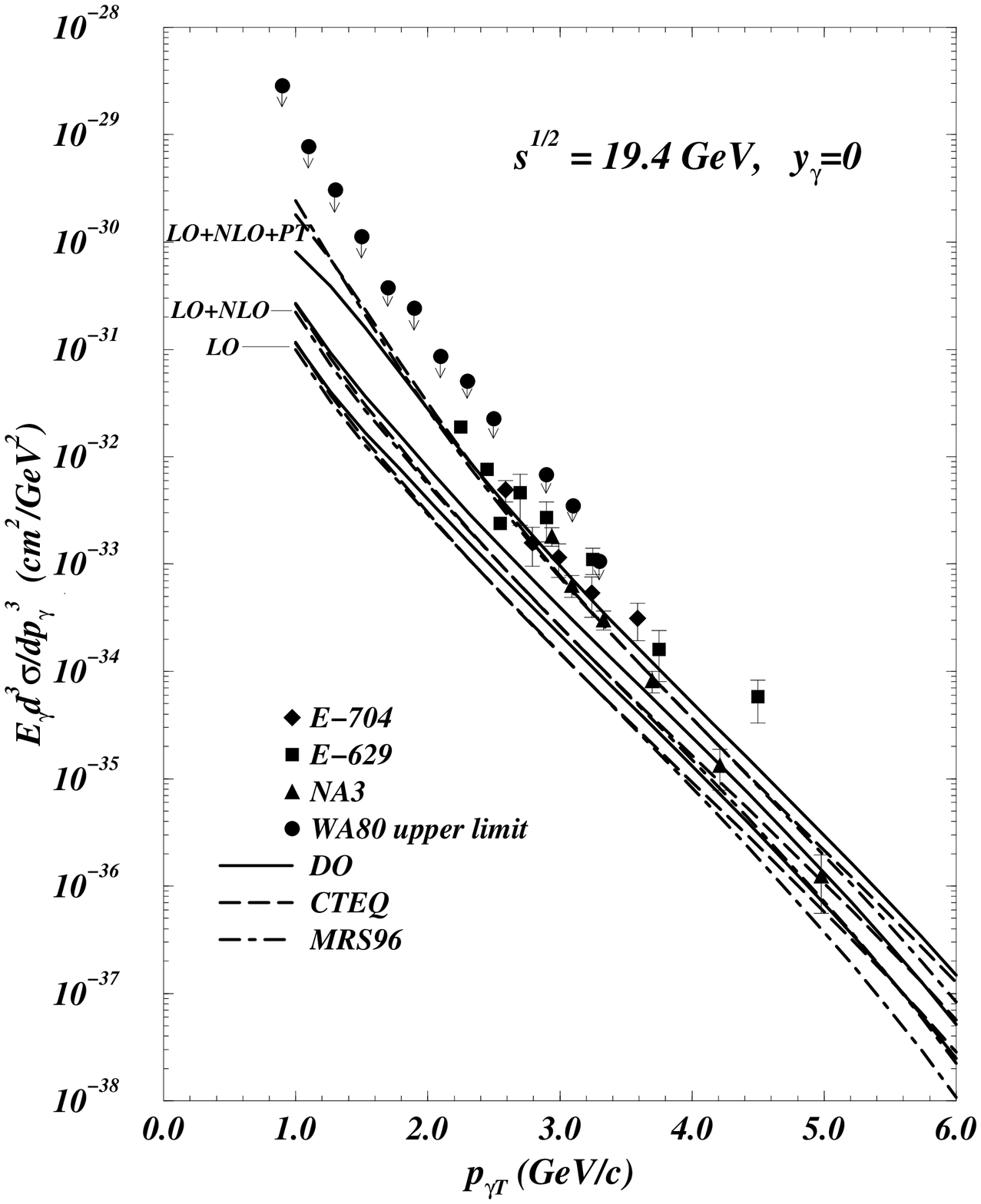}
\vskip 8.8cm
\begin{minipage}[t]{8cm}
\noindent {\bf Fig.\ 5}.  {The photon invariant cross section for
nucleon-nucleon collisions at $\sqrt{s}=19.4$ GeV and $y_\gamma=0$.  }
\end{minipage}
\vskip 4truemm

Figure 4 shows the direct photon production results for
nucleon-nucleon collisions at $\sqrt{s}$ = 63 GeV and $ y_\gamma = 0$.
The nucleon-nucleon collision data from R806/807 \cite{Ana82} and R108
(CCOR) Collaborations \cite{Ang80} are shown in the figure.  For
$p_{\gamma T} > 1.2$ GeV/c, the cross section is enhanced as $<\!
p_T^2 \!>$ increases.  The magnitude of the enhancement is smaller
than that for $\sqrt{s}$ = 19.4 GeV, indicating that the effect of
intrinsic transverse momentum decreases with increasing $\sqrt{s}$.
For $p_{\gamma T} < 1.2$ GeV/c, the cross section initially increases
as $<\! p_T^2 \!>$ increases but decreases as $<\! p_T^2 \!>$
increases further, indicating that the trend of enhancement is
reversed for $p_{\gamma T} << \sqrt{<\! p_T^2 \!>}$.  The value of
$<\! p_T^2 \!>=0.9$ (GeV/c)$^2$ provides a good description of the
experimental data.  We shall henceforth follow Owens \cite{Owe87} to
use $<\! p_T^2 \!>=0.9$ (GeV/c)$^2$ when we include the parton
intrinsic transverse momentum distribution in our subsequent
calculations.

\null\protect\vspace*{3.0cm} \epsfxsize=300pt
\includegraphics{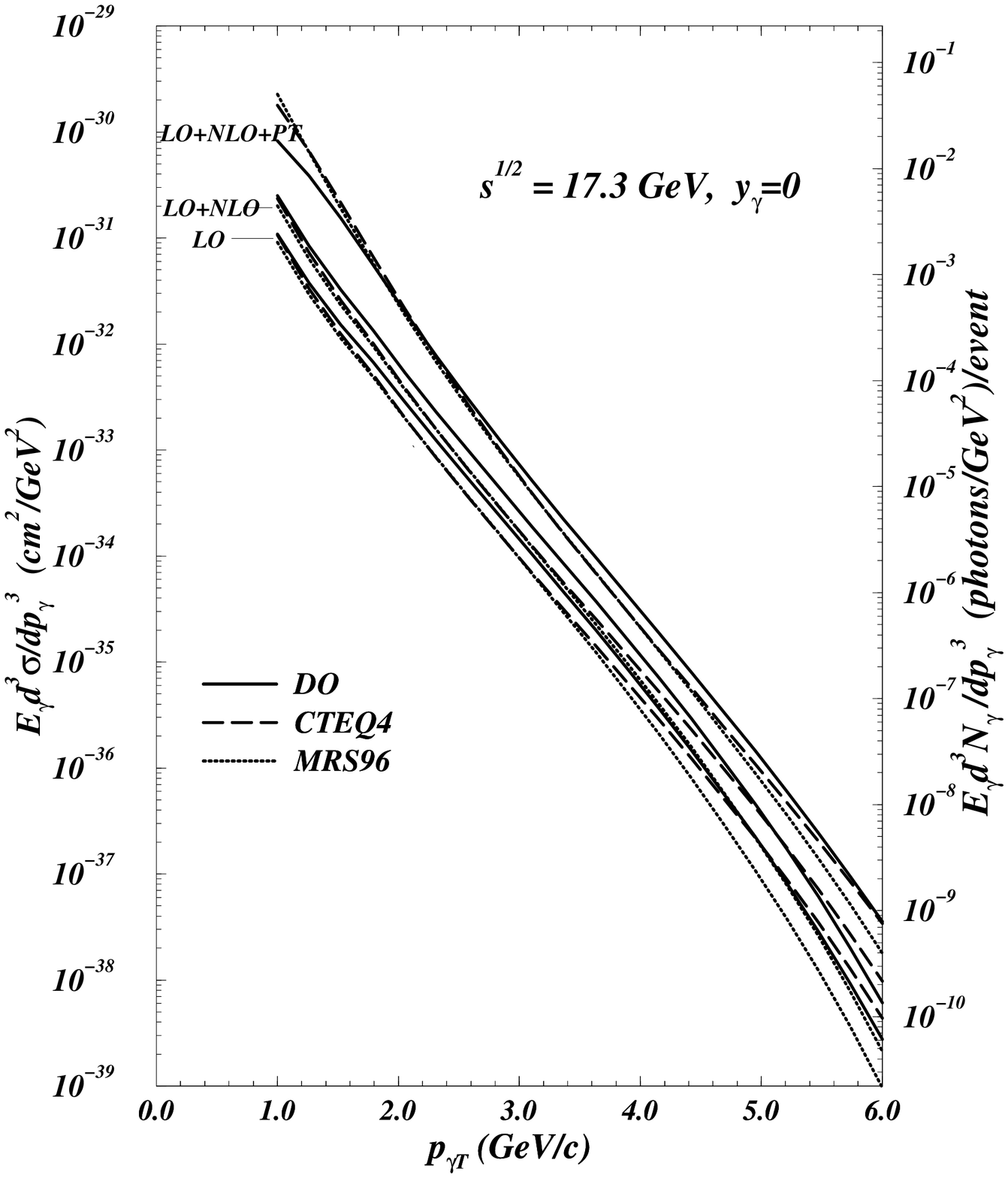}
\vskip 6.7cm
\begin{minipage}[t]{8cm}
\noindent {\bf Fig.\ 6}.  {The photon invariant cross section for
nucleon-nucleon collisions at $\sqrt{s}=17.3$ GeV and $y_\gamma=0$.
The scale on the right refers to the photon distribution per central
Pb-Pb collision event in the WA98 measurement \cite{WA98}.  }
\end{minipage}
\vskip 6truemm
\noindent 

In Fig.\ 5 we present results of the photon cross section for the $LO$
calculations, the $LO\!+\!NLO$ calculations, and the
$LO\!+\!NLO\!+\!PT$ results including further the effect of the
transverse momentum distribution, for $pp$ collisions at
$\sqrt{s}=19.4$ GeV and $y_{\gamma}=0$.  The $LO\!+\!NLO$ calculations
are enhanced from the $LO$ calculations by about a factor of 2, and
the intrinsic momentum effect brings another factor of about 4 to 8,
depending on the parton distribution function.  The intrinsic
transverse momentum of partons enhances the cross sections by a
substantial factor and is very important in the region of intermediate
$p_{\gamma T}$.  The data for $\sqrt{s}=19.4$ GeV from E704
\cite{Ada95}, E629 \cite{McL83}, and NA3 \cite{Bad86} are also shown
in Figs.\ 3 and 5.  (The NA3 data are taken from the reaction $p +
^{12}C \to \gamma + X$ where we assume a nuclear dependence of
$A^{1.0}$, as in Ref. \cite{Ada95}.)  The WA80 data in Figs.\ 3 and 5
represent the upper limits of the cross section.  The experimental
data in the region of $p_{\gamma T}$ from 2 to 4 GeV/c can be
described well when both the effects of higher-order terms and the
parton intrinsic transverse momentum are taken into account.  For the
region $p_{\gamma T}> 4$ GeV/c, the experimental data of NA3 and E629
are different.  The data of NA3 above 4 GeV/c fall within the
predictions using the CTEQ4 and the MRS96 parton distribution
functions.  Within the considerable experimental uncertainties, the
experimental data agree with the $LO\!+\!NLO\!+\!PT$ predictions,
although there are small differences in the results from different
parton distribution functions.

Recent photon production experiments using high-energy heavy ions were
performed for Pb-Pb collisions at 158 GeV per nucleon
\cite{WA98,WA9897}, which correspond to nucleon-nucleon collisions at
$\sqrt{s}=17.3$ GeV.  In order to facilitate the comparison with
experimental data, we show the theoretical results for $pp$ collisions
at $\sqrt{s}=17.3$ GeV and $y_\gamma=0$ in Fig.\ 6.  The intrinsic
transverse momentum enhances the cross sections by a factor of about 4
to 8.  It is therefore an important effect which must be included in
the discussion of photon cross sections in the region of $p_{\gamma
T}$ between 1 and 6 GeV/c.

\section{Photon Production from the QCD Hard-Scattering Processes 
in Nucleus-Nucleus Collisions}

We can determine the direct photon production cross section in
high-energy nucleus-nucleus collisions arising from the hard
scattering between partons.  Because of the weak effective shadowing
effect in photon production by parton collisions as indicated by
experimental $p$A data \cite{Zie97}, we shall neglect the shadowing
effect.  The photon distribution in a collision of a nucleus $A$ and a
nucleus $B$ at an impact parameter $\bbox {b}$ due to the hard
scattering of partons is given by
\begin{equation}
\label{eq:phm}
E_\gamma \frac{d N_{\gamma}^{A B}(H.S.)}{d \bbox{p}_\gamma}(\bbox{b}) =
n_{{}_{NN}}(\bbox{b}) 
~E_\gamma 
\frac{d N_\gamma^{NN}
(H.S.)}{d \bbox{p}_\gamma} \,,
\end{equation}
where 
$n_{{}_{NN}}(b)$ is the number of inelastic nucleon-nucleon collisions,
\begin{equation}
\label{eq:nnn}
n_{_{NN}}(\bbox{b})= A  B ~ T_{A B}(\bbox{b}) \sigma_{in}^{NN},
\end{equation}
and the photon distribution in an inelastic nucleon-nucleon collision is
\begin{equation}
\label{eq:dn}
E_\gamma \frac{d N_{\gamma}^{NN}(H.S.)}{d \bbox{p}_\gamma} = E_\gamma
 \frac{~~~d \sigma_\gamma^{NN} (H.S.)}{ \sigma_{in}^{NN}~
 d\bbox{p}_\gamma} .
\end{equation}
In the above equations, $\sigma_{in}^{NN}$ and $\sigma_\gamma^{NN}$
are the nucleon-nucleon total inelastic cross section and photon
production cross section respectively, $T_{A B}(\bbox{b})$ is the
thickness function for the nucleus-nucleus collision \cite{Won94}
\begin{equation}
T_{A B}(\bbox{b}) = \int{ d \bbox{b}_A d \bbox{b}_B T_A (\bbox{b}_A)
T_B (\bbox{b}_B) t (\bbox{b} - \bbox{b}_A - \bbox{b}_B) },
\end{equation}
where $T_A(\bbox{b})$ and $T_B(\bbox{b})$ are respectively the
thickness functions for nucleus $A$ and $B$, and the baryon-baryon
thickness function $t(\bbox{b} - \bbox{b}_A - \bbox{b}_B)$ can be
approximated by a delta function, $t(\bbox{b} - \bbox{b}_A -
\bbox{b}_B)=\delta (\bbox{b} - \bbox{b}_A - \bbox{b}_B).$ Note that
when one combines Eqs.\ (\ref{eq:phm}), (\ref{eq:nnn}), and
(\ref{eq:dn}), the nucleon-nucleon inelastic cross section
$\sigma_{in}^{NN}$ cancels out and one obtains the result
\begin{equation}
E_\gamma \frac{d N_{\gamma}^{A B}(H.S.)}{d \bbox{p}_\gamma}(\bbox{b}) =
AB T_{AB}(\bbox{b})
E_\gamma 
\frac{d \sigma_\gamma^{NN}
(H.S.)}{d \bbox{p}_\gamma} \,.
\end{equation}

Writing the result for nucleus-nucleus collisions in the form of Eq.\
(\ref{eq:phm}) provides a more intuitive picture of the scaling of the
photon multiplicity from nucleon-nucleon collisions to nucleus-nucleus
collisions.

Experimentally, one measures the yield of photons per event of
nucleus-nucleus collisions in coincidence with the measurement of the
degree of inelasticity of the events, as characterized by the
multiplicity of produced particles.  Results are given in terms of the
photon yield per inelastic collision event for the most-inelastic
fraction $f$ of all inelastic events.  The most-inelastic fraction $f$
depends on the maximum impact parameter $b_m$, and one can deduce a
relation of $f$ as a function of the maximum impact parameter $b_m$
from the Glauber model by
\begin{equation}
f(b_m)= {\int_o^{b_m} d\bbox{b} \{ 1 - (1-
T_{AB}(\bbox{b})\sigma_{in}^{NN})^{AB}
\}
\over 
\int_o^{\infty} d\bbox{b} \{ 1 - (1-
T_{AB}(\bbox{b})\sigma_{in}^{NN})^{AB}
\} }.
\end{equation}
For measurements within the most-inelastic fraction $f(b_m)$ of
events, the photon distribution per inelastic event is given by taking
the average of Eq.\ (\ref{eq:phm}) over the range of impact parameters
from 0 to $b_m$.  We therefore have the photon distribution per
inelastic event in a nucleus-nucleus collision given by
\begin{equation}
\!\!\!\!E_\gamma \frac{d N_{\gamma}^{A B}(H.S.\!)}
{d \bbox{p}_\gamma}(\bbox{b})\! =
<\! n_{{}_{NN}}\! \!\!> (b_m) E_\gamma \!\!\frac{~d \sigma_\gamma^{NN}
(H.S.\!)}{  \sigma_{in}^{NN}~ d \bbox{p}_\gamma} ,
\end{equation}
where the average number of nucleon-nucleon collisions per event
within the range of impact parameters from 0 to $b_m$ is
\begin{equation}
\!\!<\! n_{{}_{NN}} \!\!>(b_m\!)\! = \!{ {\int_0^{b_m} d\bbox{b} 
A B ~ T_{A B}(\bbox{b})
\sigma_{in}^{NN}} 
\over 
{ \int_o^{b_m}\! d\bbox{b} \{ 1\! - \!(1 \!-\!
T_{AB}(\bbox{b})\sigma_{in}^{NN})^{AB} \}} }\,.
\end{equation}

In Fig.\ 7(a) we show $<\! n{{}_{NN}} \!>(b_m$), the number of
nucleon-nucleon inelastic collisions averaged over the range
$0<b<b_m$, as a function of the maximum impact parameter $b_m$ for
Pb-Pb collisions, and in Fig.\ 7(b) the corresponding most-inelastic
fraction $f(b_m)$.

The photon production data from the WA98 Collaboration were collected
for the most-inelastic 10\% of the total number of Pb-Pb inelastic
collisions \cite{WA98}.  For this value of $f(b_m)=0.1$, the impact
parameter $b_m$ is found from Fig.\ 7 to be 4.44 fm. With an inelastic
nucleon-nucleon cross section of 29.4 mb \cite{Won94}, the average
number of inelastic nucleon-nucleon collisions within this range of
impact parameters from 0 to 4.44 fm is $<\! n{{}_{NN}} \!>=$654.  This
is the scaling number which one must multiply to the photon yield in a
nucleon-nucleon inelastic collision to give the total photon yield per
central Pb-Pb collision.

\null\protect\vspace*{3.2cm}
\epsfxsize=300pt
\includegraphics{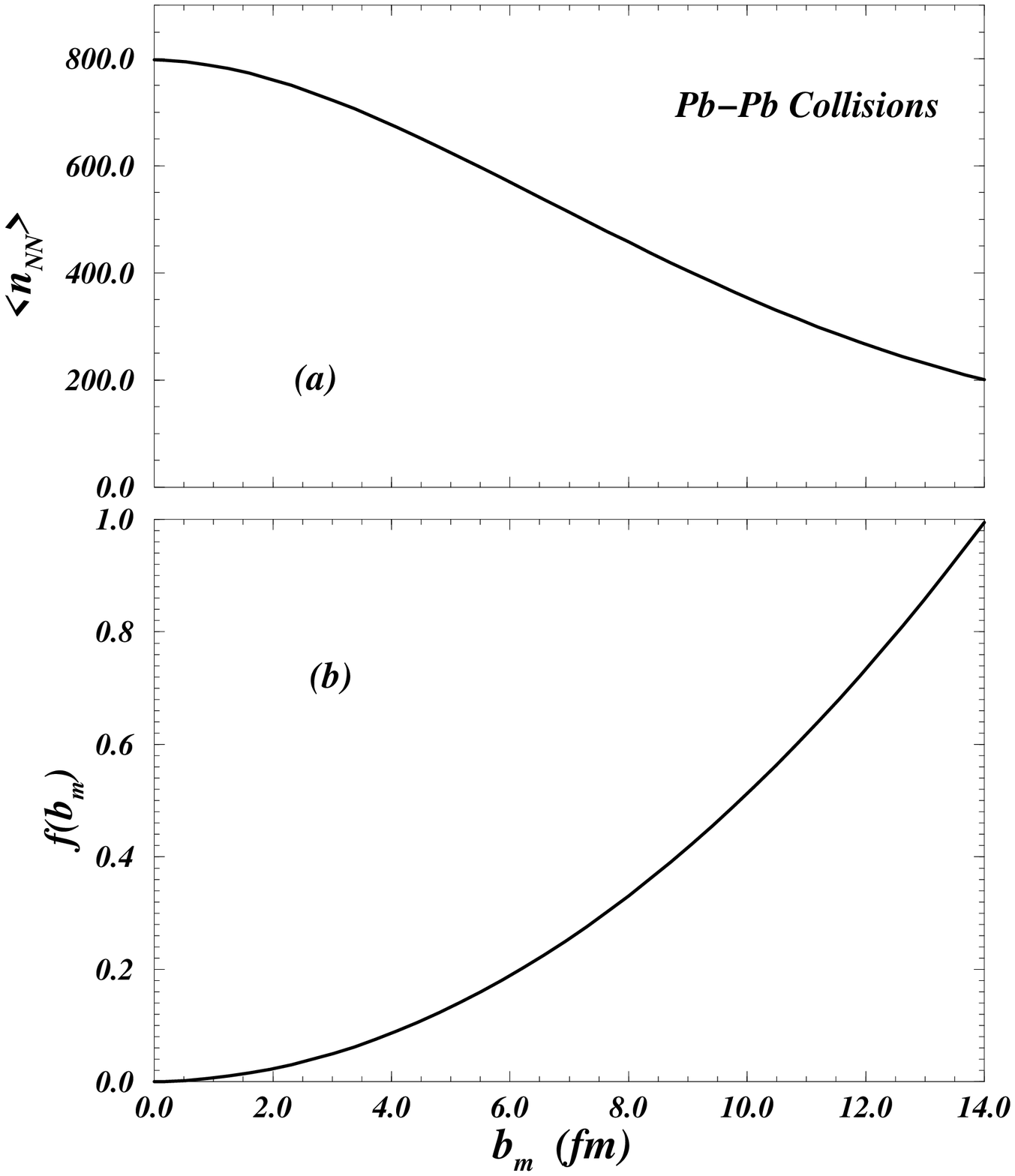}
\vspace*{6.5cm}
\begin{minipage}[t]{8cm}
\noindent {\bf Fig.\ 7}.  {(a) The number of inelastic nucleon-nucleon
collisions in Pb-Pb collisions averaged over the impact parameter from
0 to $b_m$ as a function of the impact parameter $b_m$.  (b) The
fraction of the total inelastic cross section for collisions with an
impact parameter from 0 to $b_m$.  }
\end{minipage}
\vskip 6truemm
\noindent

Using this scaling factor, we obtain the photon distribution from
hard scattering of partons per central Pb-Pb collision at a
nucleon-nucleon collision energy of $\sqrt{s}$ = 17.3 GeV and
$y_\gamma$ = 0.  They are shown as the dashed and dashed-dot curve in
Fig.\ 8 for the CTEQ4 parton distribution function (set 3) and the MRS96
parton distribution function (set 1).  They have been calculated for the
most-inelastic 10\% of the inelastic Pb-Pb collisions. They should be
compared with photon distributions from the quark-gluon plasma at
different temperatures to be discussed in the next section and shown
as the solid curves in Fig.\ 8.

The scaling between nucleon-nucleon collisions to nucleus-nucleus
collisions can be extended for S-Au collisions which were studied by
the WA80 Collaboration \cite{WA80}.  The WA80 photon data were
collected for the 7.4\% of the most-inelastic events for which the
maximum impact parameter $b_m$ is found to be 2.9 fm and the number of
nucleon-nucleon inelastic collisions $<\! n{{}_{NN}} \!>$ averaged
over the range of impact parameters from 0 to $b_m$ is 160.  This
scaling number has been used to convert the WA80 upper limit $E_\gamma
dN^{SAu}_\gamma /d^3 p_\gamma $ data to $E_\gamma
d\sigma^{NN}_\gamma/d^3 p_\gamma $ data in Figs.\ 3 and 5.

\null\protect\vspace*{2.5cm}
\epsfxsize=300pt
\includegraphics{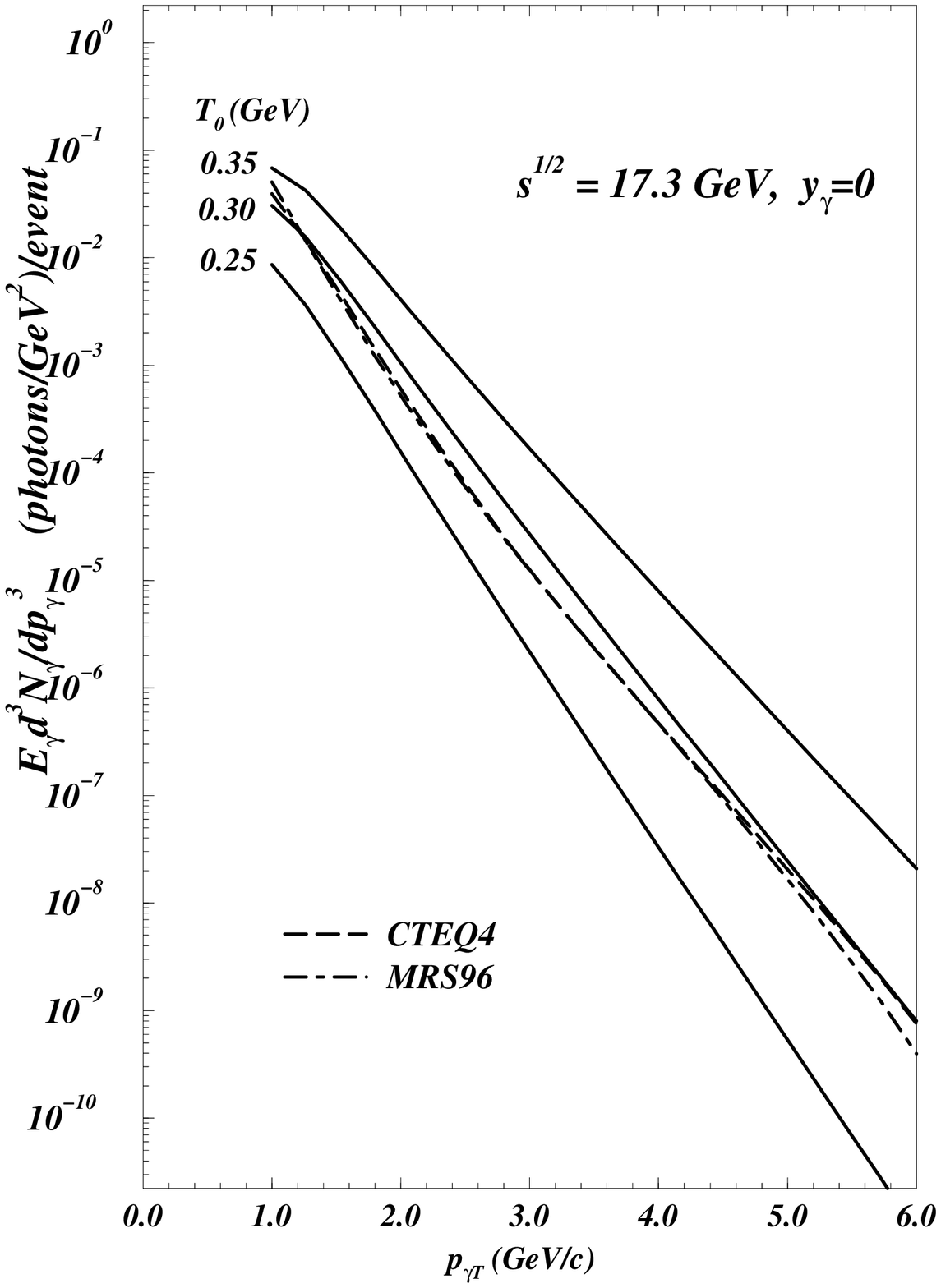}
\vskip 7.0cm
\begin{minipage}[t]{8cm}
\noindent {\bf Fig.\ 8}.  {The invariant photon distribution per
central Pb-Pb collision event in the WA98 measurement \cite{WA98} at
$y_\gamma=0$ as a function of $p_{\gamma T}$.  The dashed and
dashed-dot curves are results from parton hard scattering obtained in
$LO\!+\!NLO\!+\!PT$ calculations, and the solid curves are from a
quark-gluon plasma with different initial temperatures $T_0$ and a
critical temperature $T_c=200$ MeV.  }
\end{minipage}
\vskip 4truemm
\noindent 

\section{Photon Production from the QGP 
in Nucleus-Nucleus Collisions}

In a quark-gluon plasma, the constituents of the plasma interact with
each other.  The interaction between the constituents leads to the
production of photons via $g q ({\rm or~}\bar q)\rightarrow \gamma
q({\rm or~}\bar q)$ and $q \bar q \rightarrow \gamma g$ reactions,
which are the same processes as in photon production in parton
collisions \cite{Hwa85,McL85,Kaj86,Kap91,Sri92,Xio92}.  Upon taking
the lowest-order Feynman diagrams in the evaluation of the cross
section and assuming a thermal mass $m_{\rm th}=gT/\sqrt{6}$, the rate
of photon production in a quark-gluon plasma at temperature $T$ is
\cite{Kap91} [Eq. (16.66) of \cite{Won94}]
\begin{equation}
E_\gamma 
{ dN_\gamma \over d{\bbox{p}}_\gamma d^4 x } 
= {5 \over 9} { \alpha_e \alpha_s  \over 2 \pi^2 }
f_q(\bbox{p}_\gamma)
T^2
\ln \biggl \{ \!{ 3.7388 E_\gamma \over g^2 T } \!\biggr \}\,,
\end{equation}
where $g^2/4\pi=\alpha_s$ is the strong-interaction coupling constant,
and $f_q(\bbox{p}_\gamma)$ is the Fermi-Dirac distribution of the
quarks in the plasma which, for $E_\gamma\! >> \!T$, is approximately
the Boltzman distribution

$$f_q(\bbox{p}_\gamma) \sim e^{-E_\gamma/T}.$$ 

One can introduce the $K$-factor to take into account the
next-to-leading-order effect.  The $K(E_\gamma)$-factor as shown in
Fig.\  1(a) obtained for parton collisions (at $\sqrt{s}=$17.3 GeV) can
be used as an approximate $K$-factor for constituent collisions in the
quark-gluon plasma.  The photon distribution per event is then
\begin{equation}
\label{eq:int}
E_\gamma 
{ dN_\gamma \over d{{\bbox{p}}_\gamma } } 
= {5 \over 9} \, { K(E_\gamma)\alpha_e \alpha_s  \over 2 \pi^2 }
\int d^4x f_q(\bbox{p}_\gamma)
T^2
\ln \biggl \{ \! { 3.7388 E_\gamma \over g^2 T } \!\biggr \}\,.
\end{equation}
We shall assume Bjorken hydrodynamics in which the proper time varies
with temperature $T$ as
\begin{equation}
{\tau \over \tau_0} = \biggl (  { T_0  \over T } \biggr )^3,
\end{equation}
where $\tau_0$ is the initial proper time when the system is at a
temperature $T_0$.  We consider a system with an initial volume of $V$
given by ${\cal A} \tau_0$ where $\cal A$ is the average overlap area
for the range of impact parameters considered.  Then, during the
evolution of the system from temperature $T_0$ to the critical
phase-transition temperature $T_c$ in the quark-gluon plasma phase,
the number of photons emitted can be obtained by carrying out the
integration in Eq.\ ({\ref{eq:int}).  The result is
\begin{eqnarray} 
E_\gamma { dN_\gamma \over d{\bbox{p}}_\gamma } &=& {5
K(E_\gamma)\alpha_e \alpha_s \over 6 \pi^2 } \biggl ( { V\tau_0 T_0^4
\over \hbar^4 } \biggr ) {1 \over T_0 E_\gamma} \nonumber\\ &\times&
\biggl [ E_1(z)+e^{-z} \ln \biggl ({ 3.7388 z \over 4 \pi \alpha_s}
\biggr ) \biggr ]_{z=E_\gamma/T_c}^{z=E_\gamma/T_0} ~~,
\end{eqnarray}
where $E_1(z)$ is the exponential integral $\int_z^{\infty}dt\, e^{-t}/t$
\cite{Abr65}.

The photon distribution from the quark-gluon plasma for $T_c$=200 MeV
and different values of $T_0$ is shown in Fig.\ 8 for the 10\%
most-inelastic central collisions of Pb on Pb, where we have taken the
initial time parameter $\tau_0$ to be 1 fm/c.  As one observes, the
photon numbers increase as initial temperature increases.  This arises
because it takes a longer period of time to cool a system to the phase
transition temperature if the system is initially at a higher
temperature.  The photon number distributions from hard scattering of
partons are also shown in Fig.\ 8.  One observes that the photon
distribution from the quark-gluon plasma is greater than that from the
hard scattering when the initial temperature of the plasma exceeds
about 0.28 GeV.

\null\protect\vspace*{4.6cm}
\epsfxsize=300pt
\includegraphics{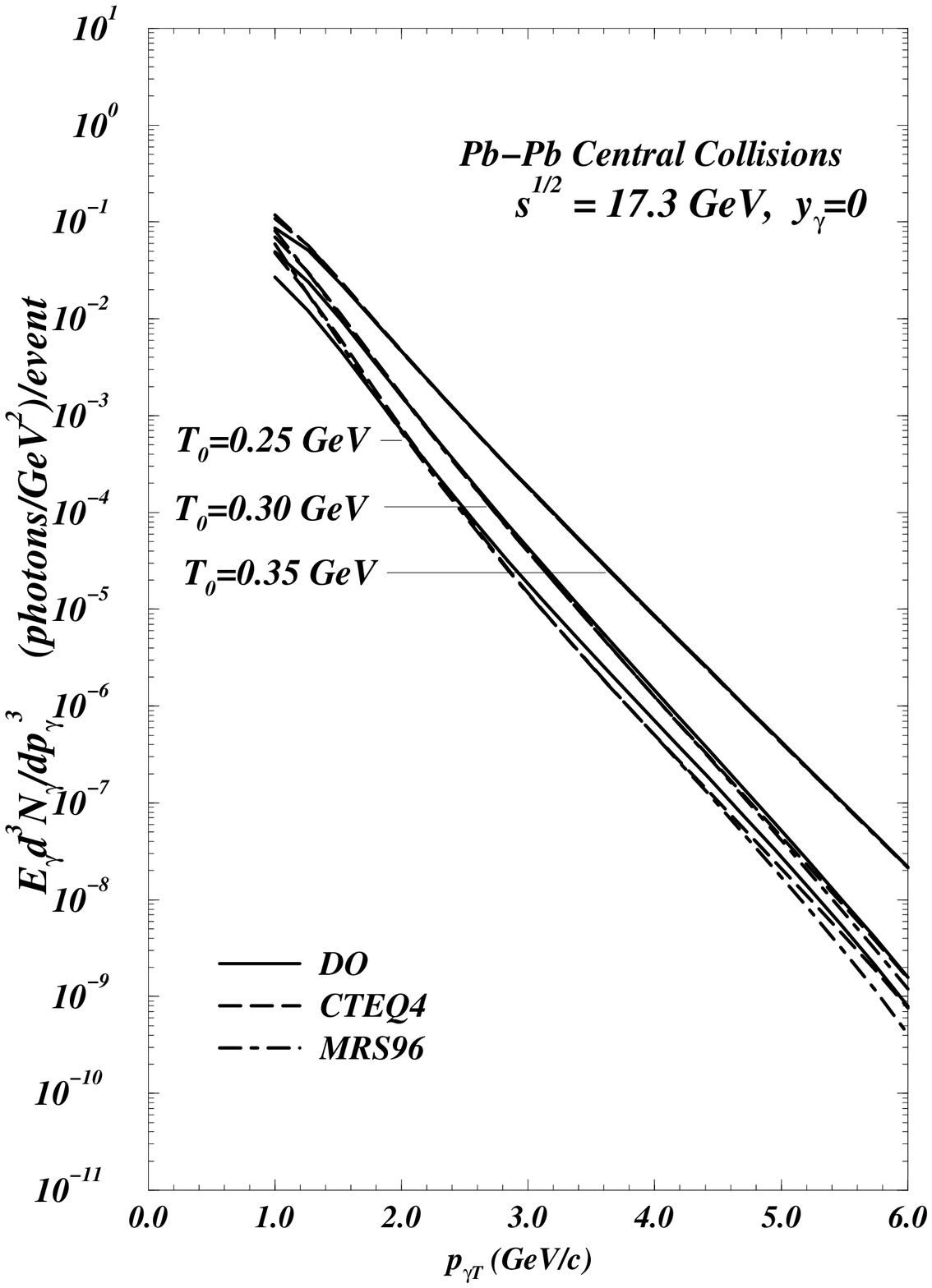}
\vspace{ 5.2cm}
\begin{minipage}[t]{8cm}
\noindent {\bf Fig.\ 9}.  {The total invariant photon distribution per
central Pb-Pb collision event in the WA98 measurement at $y_\gamma=0$
as a function of $p_{\gamma T}$, including the contributions from
nucleon-nucleon hard scatterings and the quark-gluon plasma with an
initial temperature $T_0$.   }
\end{minipage}
\vskip 4truemm

In Fig.\ 9, we show the prediction for photon production when both the
hard-scattering contributions and the quark-gluon plasma contributions
are added together, for the 10\% most-inelastic central collisions of
Pb on Pb at 158 GeV per nucleon.  The hard-scattering results are
obtained from the $LO\!+\!NLO\!+\!PT$ calculations.  The total
distribution depends on the plasma initial temperature.  The have been
calculated by assuming a critical temperature $T_c=200$ MeV.  They may
be used to compare with experimental data.

\section{Discussions and Conclusions}

In order to detect photons which are emitted during the quark-gluon
plasma phase, it is necessary to determine the contribution of photons
by non-quark-gluon plasma sources.  The hard scattering between
partons also produces photons in the momentum region of interest.  It
is therefore important to study the hard-scattering contribution to
photon production in the intermediate $p_{\gamma T}$ region of a few
GeV/c.

For photon production at $\sqrt{s} << 63$ GeV and  $p_{\gamma T}$ up
to about 6 GeV/c, the effects of the intrinsic transverse momenta of the
partons cannot be neglected.  We have examined how the intrinsic
transverse momentum distribution of partons affects the transverse
momentum distribution of the photons produced in nucleon-nucleon
collisions.  The intrinsic momentum of the partons leads to an
enhancement of the cross section by a factor ranging from 3 to 8 for
photons with intermediate $p_{\gamma T}$.  Such an effect is an
important consideration in photon measurements under investigation in
high-energy heavy-ion collisions.

We have also compared the magnitude of photons produced by hard
scattering with photons produced by the quark-gluon plasma.  We find
that for photon production in the region of 1-3 GeV/c in Pb-Pb central
collisions at 158 GeV per nucleon the quark-gluon plasma contribution
exceeds the hard-scattering contribution when the initial temperature
is greater than about 0.28 GeV.

\section{ACKNOWLEDGMENTS}
One of us (H.W.) would like to thank Prof. G. R. Young and
Prof. M. R. Strayer for their hospitality at ORNL.  The authors would
like to thank Prof. P. Aurenche and Prof. J. F. Owens for their kind
help to provide the programs for the next-to-leading-order
calculations.  The authors would like to thank Dr. T. C. Awes for
helpful discussions.  This research was supported by the Division of
Nuclear Physics, U.S. D.O.E. under Contract No. DE-AC05-96OR22464
managed by Lockheed Martin Energy Research Corp.

\end{document}